\makeatletter
\declare@file@substitution{revtex4-1.cls}{revtex4-2.cls}
\makeatother

\documentclass[twocolumn]{aastex631}

\usepackage{graphicx}
\usepackage{subcaption}
\usepackage{apjfonts}
\usepackage{amsmath, amstext}
\usepackage{hyperref}
\usepackage{listings}
\usepackage{savesym}
\savesymbol{tablenum}
\usepackage{siunitx}
\restoresymbol{SIX}{tablenum}

\usepackage{rotating}
\usepackage{xspace}
\usepackage{placeins, booktabs}
\usepackage{tikz}
\usepackage{lipsum}
\usepackage{colortbl}
\usepackage{multirow}

\usepackage{etoolbox}
\makeatletter
\patchcmd\linenumberpar{\@LN@parpgbrk}{\penalty\@LN@parpgpen\relax}{}{}
\makeatother

\newcommand{\Spider}{\textsc{Spider}\xspace}
\newcommand{\Planck}{\textit{Planck}\xspace}

\newcommand{\IRAS}{\textit{IRAS}\xspace}

\date{\today}

\begin{document}
\title{Evidence for Spatially Distinct Galactic Dust Populations}
\shorttitle{Evidence for Spatially Distinct Galactic Dust Populations}

\newcommand\Princeton{Department of Physics, Princeton University, Jadwin Hall, Princeton, NJ 08544, USA}

\newcommand\CWRU{Physics Department, Case Western Reserve University, 10900 Euclid Ave, Rockefeller Building, Cleveland, OH 44106, USA}

\newcommand\UIUCP{Department of Physics, University of Illinois at Urbana-Champaign, 1110 W. Green Street, Urbana, IL 61801, USA} 

\newcommand\UIUCA{Department of Astronomy, University of Illinois at Urbana-Champaign, 1002 W. Green Street, Urbana, IL 61801, USA}

\newcommand\Hopkins{Department of Physics and Astronomy, Johns Hopkins University, 3701 San Martin Drive, Baltimore, MD 21218 USA}

\newcommand\MPI{Max-Planck-Institute for Astronomy, Konigstuhl 17, 69117, Heidelberg, Germany}

\author{ Corwin~Shiu }
\affiliation{\Princeton}

\author{ Steven~J.~Benton }
\affiliation{\Princeton}

\author{Jeffrey~P.~Filippini}
\affiliation{\UIUCP}

\author{
Aur\'{e}lien~A.~Fraisse
}
\affiliation{\Princeton}

\author{ William~C.~Jones }
\affiliation{\Princeton} 

\author{
Johanna~M.~Nagy
}
\affiliation{\CWRU}

\author{Ivan~L.~Padilla}
\affiliation{\Hopkins}

\author{Juan~D.~Soler}
\affiliation{\MPI}

\begin{abstract}  
 We present an implementation of a Bayesian mixture model using Hamiltonian Monte Carlo (HMC) techniques to search for spatial separation of Galactic dust populations. Utilizing intensity measurements from \Planck High Frequency Instrument (HFI), we apply this model to high-latitude Galactic dust emission.
Our analysis reveals a strong preference for a spatially-varying two-population
dust model over a one-population dust model, when the latter must capture the total variance in the sky. Each dust population is well characterized by a single-component spectral energy distribution (SED) and accommodates small variations. These populations could signify two distinct components, or may originate from a one-component model with different temperatures resulting in different SED scalings.
While no spatial information is built into the likelihood, our investigation unveils large-scale spatially coherent structures with high significance, pointing to a physical origin for the observed spatial variation. These results are robust to our choice of likelihood and of input data. Furthermore, this spatially varying two-population model is the most favored from Bayesian evidence calculations. 
Incorporating \IRAS 100\,$\mu$m to constrain the Wein-side of the blackbody function, we find the dust populations differ at the $2.5\sigma$ level in the spectral index ($\beta_d$) vs. temperature $(T_d)$ plane. The presence of multiple dust populations has implications for component separation techniques frequently employed in the recovery of the cosmic microwave background.
\end{abstract}


\section{Introduction}
\label{sec:intro}

Understanding Galactic dust emission in the millimeter (mm) wavelength range plays an important role in cosmic microwave background (CMB) science. This diffuse emission, originating from $\sim$20\,K  thermal emission of interstellar dust grains within our Milky Way galaxy, peaks at $\sim$2\,THz, but remains significant relative to CMB anisotropy at frequencies above about 100\,GHz. Understanding and characterizing Galactic dust emission is essential for accurately modeling its emission to correct for its distortions of the underlying CMB signal.

The properties of interstellar dust grains, including their size and shape, composition, distribution, and temperature, all influence the resulting radiation \citep{draineli_2001, draine_2007, draine_2009}. In the context of component separation, the complexity of interstellar dust is typically flattened; its spectral energy distribution (SED) is modeled as a modified blackbody, with emissions resembling that of a thermal blackbody at temperature $T_d$ and a fractional emissivity that scales with frequency according to its spectral index $\beta_d$. \citep{brandt1994,Finkbeiner_1999}.

Much of our understanding of the mm-wave emission of dust comes from two satellite data sets: first from \IRAS \citep{iras} and then \Planck \citep{planck_I_2018, planck_III_2018}.  

\Planck uses three bands, 353, 545 and 857\,GHz, in combination with DIRBE \IRAS 100\,$\mu$m to construct an all-sky model of dust-emission parameterized by a single-component emissivity spectral index and temperature $(\beta_d,T_d)$ that vary along the line-of-sight \citep{planck_XI_2013, planck_XVII_2014}. 

Meisner and Finkbeiner observed that there is a flattening of the SED model at lower frequencies (100-350\,GHz) that departs from a single-component SED. Instead, they proposed a 
two-component model with spatially fixed spectral parameters that incorporates the superposition of a hot and cold component to more accurately capture the shape of the dust SED at these lower frequencies \citep{Finkbeiner_1999,Meisner_2014}. 

Furthermore, Hensley and Draine have shown that dust grains with distinct sizes and compositions will attain different temperatures, even in the same radiative environment. They have developed a physically motivated two-component dust model positing that dust emission originates from a mixture of carbonaceous and silicate sources \citep{Hensley_2018, Hensley_2021}.
In their recent work, however, Hensley and Draine state that the observed lack of frequency dependence in the far-infrared polarization fraction is most naturally explained by a single-composition dust model \citep{hensley2022astrodustpah}.

In a different approach, \cite{Delouis_2021} utilizes the stability of the solar dipole to construct a model for SED spatial variations of the interstellar dust emission. They model dust with perturbative expansions of the spectral index and establish large-scale spectral variations.

Given the inherent complexity of Galactic dust emission, there exists considerable scope for refining modeling efforts. 
In this paper, we exclusively examine dust emission in intensity and present an approach that looks for SED variations in the spatial domain. We identify variations of the SED and assign them to distinct global populations of dust. 
Here, we delineate between a ``population'', a modeled concept, and a ``component'', a physical entity that is defined by its chemical composition, such as the relative abundance of silicon and carbon. While multiple populations may arise due to the presence of multiple dust components, they can also arise from a single component experiencing differing temperatures, changing the SED. 
In this analysis, we are unable to disentangle these compensatory effects on the SED.


In Section \ref{sec:regression}, we first describe a generic method for linear regression with uncertainties. Then, in Section \ref{sec:gaussian_mix_model}, we utilize this method to build a likelihood model for dust emission from multiple dust populations. In Section \ref{sec:sims}, we demonstrate that this likelihood is unbiased and can recover components even in highly mixed situations. In Section \ref{sec:preprocess}, we describe the preprocessing steps for this analysis, and Section \ref{sec:lr40_analysis} provides the primary result of this analysis. In Section \ref{sec:spider_analysis}, we look for the same features in the region observed by the \Spider balloon experiment \citep{filippini_spider, Rahlin2014, Gualtieri2018LTD}, a relatively large ($\sim 5\%$) and clean patch of the sky used for CMB B-mode studies. Then, in Section \ref{sec:analysis_choices} explore several analysis choices and how these choices affect the results. In Section \ref{sec:evidence}, we calculate the Bayesian evidence to determine which model is most favored. Lastly, in Section \ref{sec:inferred_dust}, we fit a modified blackbody model (MBB) to the recovered populations. 

\section{Method} 
\label{sec:method}
\subsection{``Mahalanobis'' regression in the generic case with X,Y uncertainties}
\label{sec:regression}
\begin{figure}[h]
\centering
\begin{tikzpicture} 
\draw[->, thick](0,0) to (0, 5) node[left]{$y$}; 
\draw[->, thick](0,0) to (5,0) node[below]{$x$};
\node(point) at (1.5,4){};
\draw(point)[color = blue!10, fill = blue!10]circle[x radius = 1em, y radius = 4em, rotate = 0];
\draw(point)[color = blue!30, fill = blue!30]circle[x radius = 0.5em, y radius = 2em, rotate = 0];
\draw(point) [color = black, fill = black]circle(.4ex);
\draw (point) node[above]{${\bf Z_i}$};
\draw (point) [dashed, black] to (3.0, 4); 
\draw(point)[dashed, black] to (0.5, 4); 
\draw[blue](1.85,4) node[thick, font=\bfseries\large, below]{$\phi$}; 
\draw[thick, blue, <->](2.25, 4) arc(0:-80:0.75); 
\draw(2.5, 3.75) node[right, font=\small]{$\frac{\pi}{2} - \theta$};
\draw[<->](2.5, 4) arc(0:-63.4:1.0);
\draw[<->, thick, red](-1.5,  0)to (0, 0.75)to (4.5, 3.0);  
\draw[red](0,0) -- (0, 0.75);
\draw[red](0, 0.375) node[left]{$b$}; 
\draw[red](0.7, 0.92) node[]{$\theta$};
\draw[dashed, thin, black](0,0.75) -- (1.5, 0.75); 
\draw[ black](point) to (2.15, 2.7) node[right]{$\Delta_i$} to(2.5, 2);
\draw[black, thin](2.7, 2.1) to (2.6, 2.3) to (2.4, 2.2);
\draw[blue, thick](point) to (2, 1.75);
\draw[blue, ->](point) to(1.6, 3.55)node[left ]{${\bf u}$};
\draw[->, black](0, 0.75) to (-0.25, 1.25) node[left]{${\bf v}$};
\draw[black, thin](-0.1, 0.95) to (0.1, 1.05) to(0.2, 0.85); 
\end{tikzpicture}
\caption{Diagram of fitting a point ${\bf Z_i}$ with covariance ${\bf \Sigma_i}$ to a model line $(\theta, b)$. A vector ${\bf v} $ perpendicular to the slope $\theta$ is constructed so that we can project the point ${\bf Z_i}$ onto the line, ensuring that the perpendicular distance is given by $\Delta_i = {\bf v}^\mathrm{T} {\bf Z_i}  -b \cos \theta$. While this may be the minimum geometric distance, the minimum Mahalanobis distance is along the vector ${\bf u_i} = (\cos \phi, -\sin \phi)$. Then, the uncertainties $\Sigma_i$ are projected along the vector ${\bf u_i}$ to the line: $S_i = {\bf u^\mathrm{T}_i \Sigma_i  u_i}$. }
\label{fig:orth_reg} 
\end{figure}
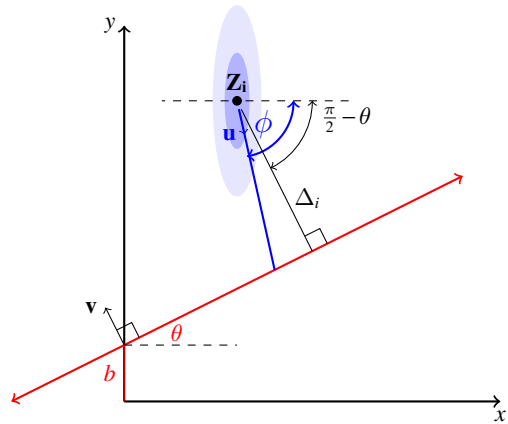

Traditional linear regression assumes that independent variables (the abscissa) are known with certainty. In many real-world scenarios, however, uncertainties and errors in data collection can affect both the abscissa and ordinate. Indeed, in our analysis of the \Planck data, the uncertainties are of a similar order to each other.
Ignoring or underestimating these uncertainties when using standard methods, such as ordinary least squares, can lead to biased parameter estimates.

Handling "Errors-in-variables" in linear regression is a challenging topic. Unlike the standard approach, incorporating uncertainties in both the abscissa and ordinate requires sophisticated methods to account for uncertainty propagation. Moreover, there is no consensus on the best approach \citep{hogg_likelihood, tullyfisher_fitting}.

Orthogonal regression is a popular approach that minimizes the orthogonal distance of all the points to a regression line. While the orthogonal distance is the smallest geometric distance, however, it may not correspond to the smallest distance in probability space in cases where errors in the abscissa and ordinate are unequal. We make a modification to the method to minimize the Mahalanobis distance between each point $i$ and the regression line $k$. Therefore, we find the distance along a unit-vector ${\bf u_i}$ whose direction is described by the angle $\phi$,

\begin{align}
\phi = \arctan\left( -\frac{1}{\tan \theta}\frac{\sigma_y^2}{\sigma_x^2} \right),
\label{eq:ortho_angle}
\end{align}
and the uncertainties of each point $i$ is also projected along ${\bf u_i}$. The derivation of this expression can be found in Appendix \ref{sec:appendix_derivation}.
We call this method ``Mahalanobis regression''. 

This regression method, similar to orthogonal regression, fails to capture the variance along the regression line.
Handling the full uncertainty in the generic case is computationally challenging.
Rather than reducing dimensionality by using a regression model that has two variables, namely a slope and an offset, a fully generic regression would require $N + 2$ variables to account for $N$ data points. Each observed point represents a sample from a two-dimensional Gaussian distribution, originating from a pair of ``true'' values ($x^*, y^*$) within a regression model.

However, this approach becomes computationally impractical when considering the construction of a mixture model in the following section. Therefore, we chose to reduce the dimensionality of the problem using Mahalanobis regression as described.

\subsection{Gaussian Mixture Model}
\label{sec:gaussian_mix_model}
We introduce a generic Gaussian Mixture Model (GMM), in which a given map pixel $i$ can belong to one of $K$ populations. Depending on the specifics of the analysis, certain parameters can be constrained to have a fixed value of zero. 
We parameterize each population by its slope $\theta_{k}$, calibration offset $b_k$, and assign each point $i$ a population using $q_{i,k}$. $q_{i,k}$ is a Boolean value for point $i$ to belong in population $k$. We define $q_{i,k} = 0$ as a rejection of point $i$ from population $k$, forcing that point to be assigned to a different population. 

We fit this model to a scatter plot of points using \emph{Mahalanobis regression}, as defined in the previous section, performed by maximizing the following likelihood:
\begin{align}
  \mathcal{L} ( \{q_{i,k} \} &, \theta_k, b_k, V_k, p_k) =  \nonumber \\ &\begin{aligned}  
  \prod_i \prod_k&\left[ \frac{1}{\sqrt{2 \pi (\Sigma_{i,k} + V_k)}} \right. \\ 
  &\times \left. \exp \left( - \frac{(\Delta_{i,k}/\cos(|\phi| + \theta - \pi/2) )^2}{2( \Sigma_{i,k} + V_k)} \right) \right]^{ q_{i,k}} ,
  \label{eq:Kparamdust} 
  \end{aligned} \\
  \Delta_{i,k} &=  {\bf v} \left(\theta_{k}\right)^\mathrm{T}  Z_i - b_k\cos(\theta_k) = -\sin\left(\theta_k \right) x_i + \cos\left(\theta_k\right)(y_i - b_k), \label{eq:ortho_distance} \\
  \Sigma_{i,k} &={\bf u_i} \left(\theta_k \right)^\mathrm{T} \begin{pmatrix} \sigma_x^2 & 0 \\0 & \sigma_y^2 \end{pmatrix} {\bf u_i}\left(\theta_k \right) = \sigma_x^2 \cos^2 \phi + \sigma_y^2 \sin^2 \phi . \label{eq:proj_sigma}
\end{align}
The orthogonal distance of point $i$ to line $k$ is represented by $\Delta_{i,k}$. Then, the Mahalanobis distance can be computed by simple trigonometry: $\Delta_{i,k}/\cos(|\phi| + \theta - \pi/2)$. To get the uncertainties, we use Eq. \ref{eq:proj_sigma} to project the covariance of each data point along the Mahalanobis distance, ${\bf u}$, resulting in the scalar value $\Sigma_{i,k}$. Furthermore, we assign an additional variance term $V_k$ to each population, representing the intrinsic scatter fitted for that population. 

We further impose a prior on $q_i$ of the form of a Dirichlet-Multinomial distribution, 
\begin{equation}
  q_{i,k} \sim \text{DirMult}(n = 1, \{ p_A, p_B, \cdots p_K \} ).
  \label{eq:betabin}
\end{equation}

\noindent The hyperparameters $\{p_A \cdots p_K \}$ have an additional constraint that $\sum_k p_k= 1$ so that the prior distribution therefore has an expected value of $p_k$ for each population $k$. This has a physically motivated interpretation that $p_k$ represents the global fraction of points belonging to the $k^{\mathrm{th}}$ population. 

In the common case where there are only two populations, this prior distribution reduces to the beta-binomial distribution. Drawing from this prior can be thought of as a two-step process: first sample from a Beta($\alpha = p_A, \beta \ = 1 - p_A$) to get a probability $\pi_{i,A}$ for each point, then run a Bernoulli trial with probability $\{ \pi_{i,A}, (1 - \pi_{i,A})\}$ to determine the assignment $q_{i}$ of this point. 
With multiple populations, this can be conceptually generalized to drawing a probability from a Dirichlet distribution and then drawing from a categorical distribution.  

We treat $\{p_A \cdots p_K \}$ as global hyper-parameters, and fit for them simultaneously alongside the other parameters. This likelihood construction limits the propagation of noise into our final population assignments: statistically indistinct points will largely be determined by the prior with $\pi_{i,A} = p_A$. Effectively, these points are assigned to a population based on prior beliefs about overall prevalences.

\subsection{Implementation in Python}

Scikit-learn is perhaps the most well-established and widely used open-source implementation of GMMs \citep{scikit-learn}. The code is based on the Expectation-Maximization (EM) algorithm \citep{em_paper}. While this algorithm is computationally efficient and converges far faster than Monte Carlo methods, EM methods provide only a point estimate of the MLE. Monte Carlo methods, on the other hand, naturally estimate uncertainty by sampling the posterior distribution and are therefore applied in this analysis. 

Additionally, we need estimates of $\pi_{i,k}$, the probability a given pixel belongs to a population $k$. 
Therefore, we need to sample the full likelihood (Eq. \ref{eq:Kparamdust}), \, including the discrete variables, $q_{i,k}$, which is a task not handled by all probabilistic programming packages \citep{stan}. 
This analysis is computationally challenging, because $N$ is the number of pixels in our region of interest, of order $10^5$, and the probability volume is $\sim K^N$, representing an enormous space for MCMCs. 
The computational cost of any Monte-Carlo algorithm can be expressed as $\mathcal{O}(M*N)$ where $M$ is the number of Monte-Carlo samples and $N$ is the dimensionality. This is because proposing new states scales linearly with the number of dimensions. 
However, each proposal step is not independent; for a random-walker algorithm, $M \sim \mathcal{O}(N)$ generates an independent point \citep{neal_hmc}. 
Therefore, a Metropolis-Hastings algorithm is expected to scale as $\mathcal{O}(N^2)$ making it computationally impossible to use rejection samplers in high-dimensional spaces. 

Hamiltonian Monte-Carlo (HMC) instead uses dynamical properties to more efficiently generate independent points \citep{hmc}. The proposals are more effective for exploring high-dimensional spaces, making them particularly useful for complex Bayesian inference. Independent points scale according to $M \sim N^{\frac{1}{4}}$ \citep{neal_hmc, beskos2010optimal}, so the total algorithm scales as $\mathcal{O}(N^{\frac{5}{4}})$. 

This analysis leverages \emph{NumPyro}, which is a probabilistic programming language that relies on \emph{JAX} for automatic differentiation of functions \citep{phan2019composable, bingham2019pyro} and efficiently generates HMC samples.

\section{Validations on Simulations} 
\label{sec:sims}
\begin{figure*}
\centering
\includegraphics[width = 0.8\textwidth]{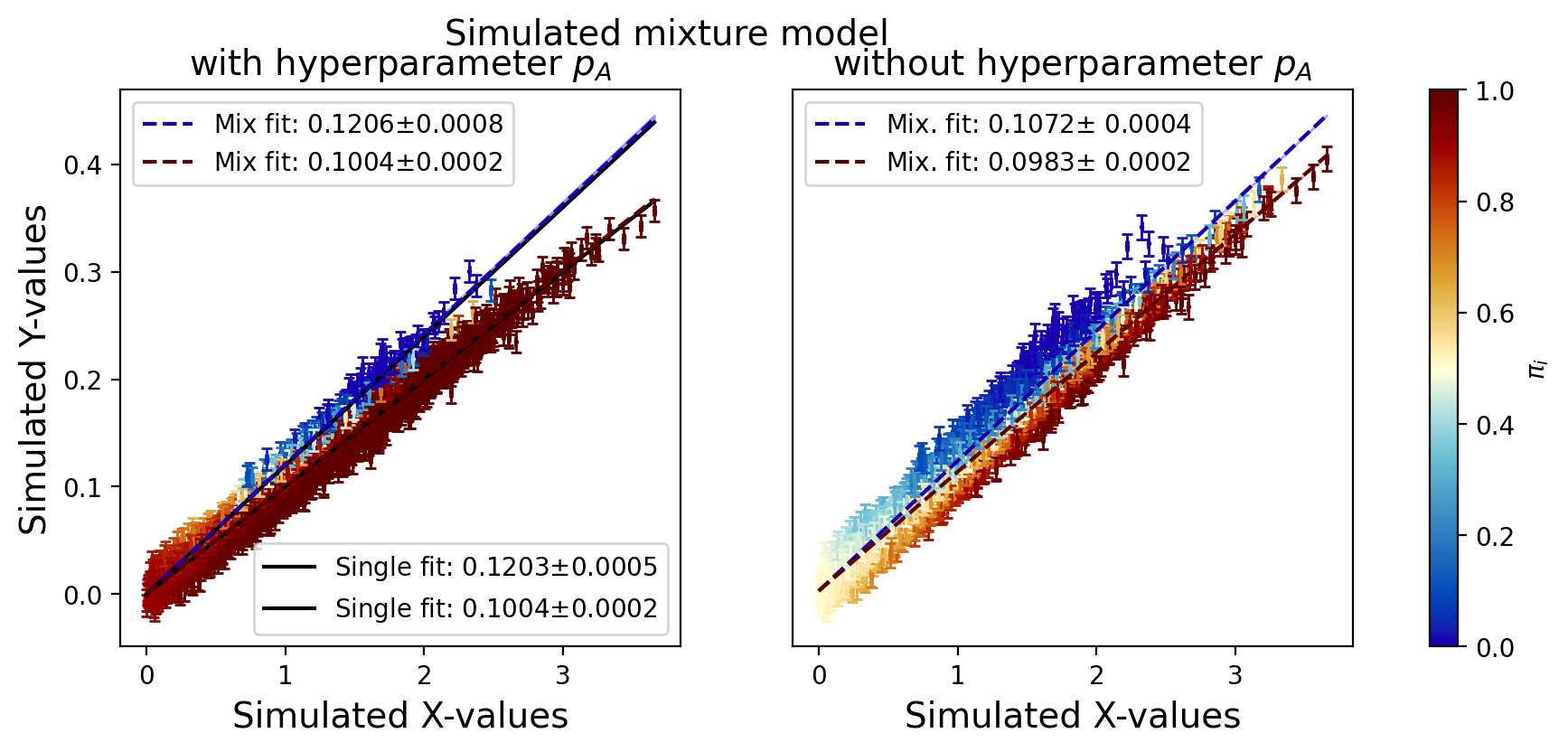}
\caption{We conducted two validation analyses using Mahalanobis regression (Eq. \ref{eq:Kparamdust}) on the same input dataset. On the left, we assumed two distinct populations ($m_0, m_1$) and estimated a global population fraction parameter $(p_A)$ without including offset or additional variance parameters $(b_k = 0, V_k = 0)$. On the right, the hyperparameter $p_A$ is turned off and the prior probability of each point $\pi_i$ is instead drawn from a uniform distribution. These fits are labeled ``Mix. fit'' for the two populations respectively. These are compared to ``Single fit,'' which is the Mahalanobis regression for each of the input populations. The colors represent the probability of a data point $p_i$ belonging to population 1.  On the left, even with two highly overlapping populations, we recover the inputs and can distinguish points to their respective population with high confidence. On the right, without this global hyperparameter, points with an equivalent likelihood of belonging to either population return $\pi_i \sim 0.5$. Because the steeper blue population intrinsically has a far smaller prevalence, this equal assignment biases the slope downward, and we fail to recover the input slopes.} 
\label{fig:nuts_example}
\end{figure*}

To validate this analysis method, we can generate data consisting of two populations with similar slopes and show that the pipeline is capable of distinguishing the populations. 
Figure \ref{fig:nuts_example} shows an example of such a test, where two populations with very different proportions $(N_{\{0,1\}}= \{200,2000\})$ and input slopes $(\theta_{\{0, 1\}} = \{0.10, 0.12\})$ are observed with highly overlapping populations. In this simulated analysis, we do not include an offset $b_k = 0$ or variance $V_k = 0$ in the model. The figures then differ on whether they include the fitted hyperparameter $p_k$ in the analysis. 

On the left, we use the nominally described method, and we are able to accurately distinguish points that belong to each population in the extremes with high probability $\pi_i$. 
Points clustered in the middle, well within the uncertainties for either model $\theta_{\{0,1\}}$, are preferentially assigned to the population that is more numerous. 
The fitted hyperparameter $p_A$ can be interpreted as the global population fraction. In the case where a point's uncertainty makes it indistinguishable from both populations, $p_A$ acts as a prior on how to assign that point through the $\pi_i$'s.

On the right, we disable the hyperparameter, $p_k$, and instead draw $\pi_i$ from a uniform prior ranging from 0 to 1. In this case, points clustered between the populations are assigned $\pi_i \approx 0.5$ indicating no preference for either population. However, due to the difference in prevalence of the two populations, there is an over-assignment of points to the blue population, introducing a bias in the recovered slope. 

While a uniform prior may seem impartial at the individual data point level, its effects are less innocent on a global scale. A uniform distribution is equivalent to a Beta distribution with fixed $\alpha = \beta = 1$, resulting in a mean of 0.5. The presence of this uniform prior is globally informative, leading to incorrect assignments of populations and ultimately causing a failure in the recovery of the simulated inputs. 

For this reason, we include the fitted hyperparameter $p_k$ in all configurations when running the data analysis.

\section{Preprocessing Pipeline}
\label{sec:preprocess}

We begin with the complete set of full-mission \Planck HFI total intensity maps: 100, 143, 217, 353, 545, and 857\,GHz\footnote{\Planck public data release 3.01. In this document, all \lstinline{.fits} files are found on the \Planck Legacy Archive}. 
The latter two HFI maps are converted from intensity units (MJy/sr) to thermodynamic units ($\mu\mathrm{K_{cmb}}$). 
All maps are then smoothed to a common 10-arcminute resolution. 
Next, the cosmic microwave background anisotropies are subtracted to get a clean measurement of dust. For this, we use the \Planck 2018 component-separated CMB map from their SMICA pipeline~\footnote{COM\_CMB\_IQU-smica\_2048\_R3.00\_full.fits }.
All maps are converted back to MJy/sr units following the \IRAS convention, assuming a SED $I_\nu \propto 1/\nu$.

\begin{table}
\captionsetup{size = footnotesize}
\caption{Key numbers for the preprocessing pipeline in this analysis. The entries in the first row are re-derived unit conversions used to convert thermodynamic units ($\mathrm{K_{cmb}}$) to flux density following the \IRAS convention of a $I_\nu \propto \nu^{-1}$. RIMO bandpasses were used in this calculation. All values in this row are consistent with \Planck's published values to less than half a percentage \citep{planck_XI_2013}. The second row shows the average noise level, computed from an ensemble of FFP10 simulations, converted to MJy/sr over the cleanest 40\% of sky. } 
\begin{tabular*}{\columnwidth}{@{\extracolsep{\fill}}p{2.0cm}@{\hskip 0.2cm}c@{\hskip 0.2cm}c@{\hskip 0.2cm}c@{\hskip 0.2cm}c@{\hskip 0.2cm}c@{\hskip 0.2cm}c}
\toprule
Planck Band & 100 & 143 & 217 & 353 & 545 & 857 \\ \midrule
\hbox{Unit Conversion} (MJy/sr $\mathrm{K_{cmb}}^{-1}$) & 243.5 & 371.1 & 482.4 & 288.1 & 58.02 & 2.29 \\
\hbox{Avg. Noise level} (MJy/sr) $\times 10^3$ & 0.7 & 1.0 & 2.3 & 1.5 & 2.0 & 1.9\\ \bottomrule
\end{tabular*}
\label{table:unit_conv}
\end{table}

Noise estimates are needed per-pixel for this analysis. For the Planck dataset, we use an ensemble of FFP10 simulations, pixelized and converted to MJy/sr units. 
We found that the half-mission difference maps have more variance than expected from FFP10 simulations for the Planck sub-mm channels: 545\,GHz and 857\,GHz. In the interest of making a conservative set of assumptions, the noise estimates used for these frequencies are scaled up by 50\%, which is a scaling more than necessary to account for the observed variance. Because the signal-to-noise ratio is so high in these channels, the conclusions of this analysis are not sensitive to this addition of noise. 

The subtraction of the SMICA CMB template introduces additional noise. We can estimate the inherent noise of the CMB template by looking at the variance of the half-mission difference map and scaling by a factor of four\footnote{a factor of two due to data splits, and another factor of two due to considering the distribution of the difference of two Normals.}. For the cleanest 40\% of the sky and at $N_{\mathrm{side} }= 64$, the half-mission difference variance is approximately $\sim 4.4\,\mu \mathrm{K_{cmb}}^2$. 
 Therefore, the estimated pixel noise comes from the quadrature addition of FFP10 simulations with this estimated CMB template noise term. This factor only increases the uncertainty of \Planck 217\,GHz by 5\% and \Planck 353\,GHz by 1\%. At 545\,GHz and 857\,GHz, the CMB template noise addition term is negligible.

Zero-level offsets must be applied to all these dust maps to avoid monopole artifacts in scaling. A typical approach is to use HI as a tracer of dust and to require that when HI trends to zero, the dust emission is also expected to be zero \citep{planck_XI_2013}. In the interest of making a minimal set of assumptions, we can circumvent this issue by including this monopole offset in our fits.

We then mask the resulting maps to different regions and excise point sources\footnote{\lstinline{HFI_Mask_PointSrc_2048_R2.00.fits}}. Lastly, we pixelate the maps from their native $N_{\mathrm{side}}$ to $N_{\mathrm{side}} = 32,64,128$, which effectively applies a $\sim$ 110, 55, and 27 arcmin beam to these maps, respectively.  We have both computational and astrophysical motivations for this. From an astrophysical perspective, we encounter the most problematic foregrounds for CMB component separation at the largest angular scales. Furthermore, we can neglect the cosmic infrared background (CIB) at these angular scales, as confirmed by appropriately masking and pixelating derived CIB templates \citep{Lenz_2019}. 

We neglect other sources of foregrounds in intensity in this analysis. We do not model free-free and synchrotron emission, as they are expected to be substantially less significant at the mm and sub-mm wavelengths \citep{planck_I_2018}. Furthermore, CO emission is primarily contained in the Galactic plane, and we anticipate that the contamination in our high-latitude region of the sky will be minimal \citep{planck_XIII_2013}.

\section{Global Dust Populations over the Cleanest 40\% Sky Region}
\label{sec:lr40_analysis}
\begin{figure*} 
\centering
%
\includegraphics[width=\textwidth]{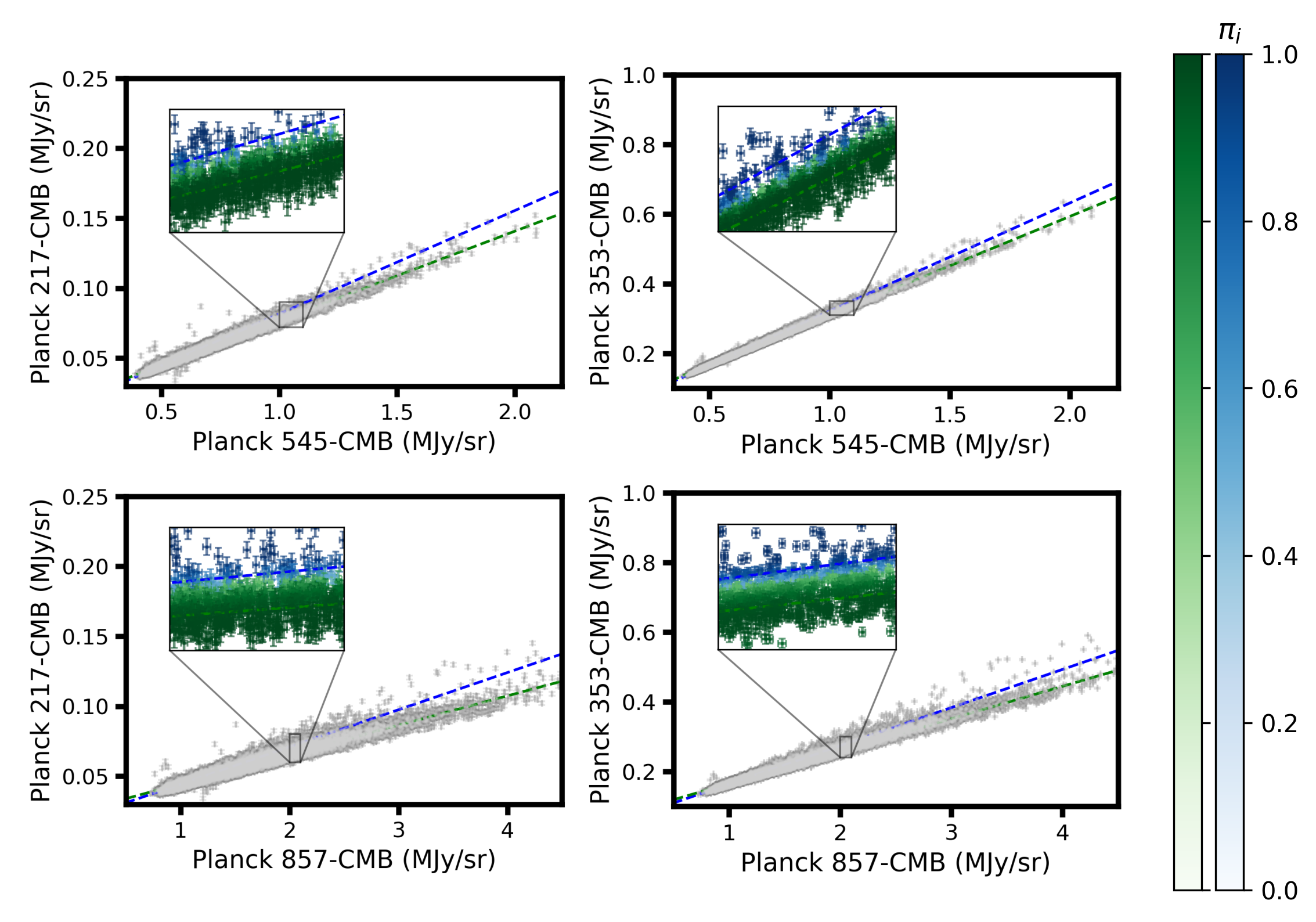}
\includegraphics[width = 0.24\textwidth]{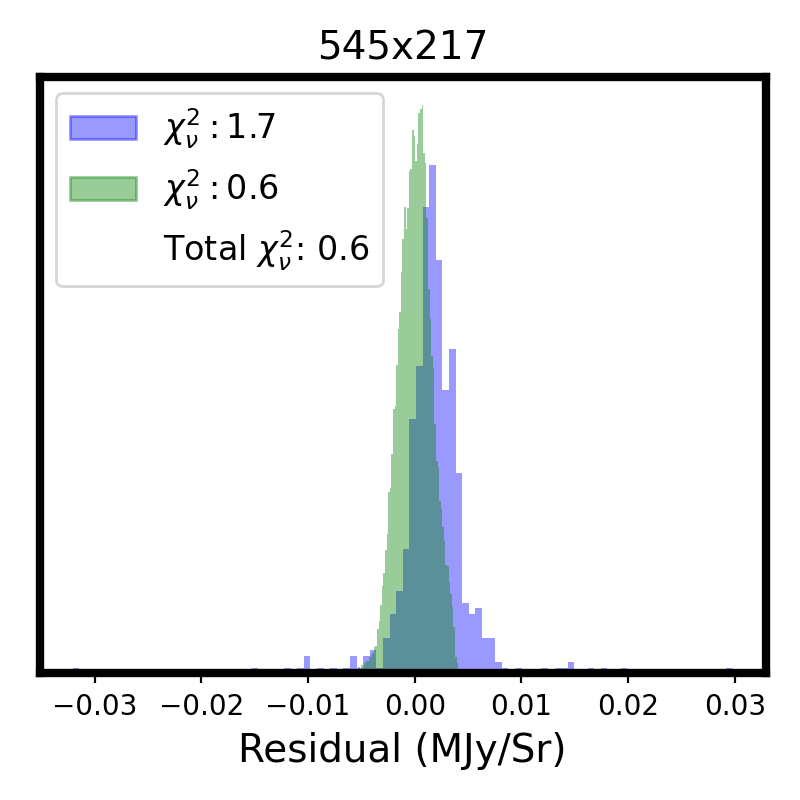}
\includegraphics[width = 0.24\textwidth]{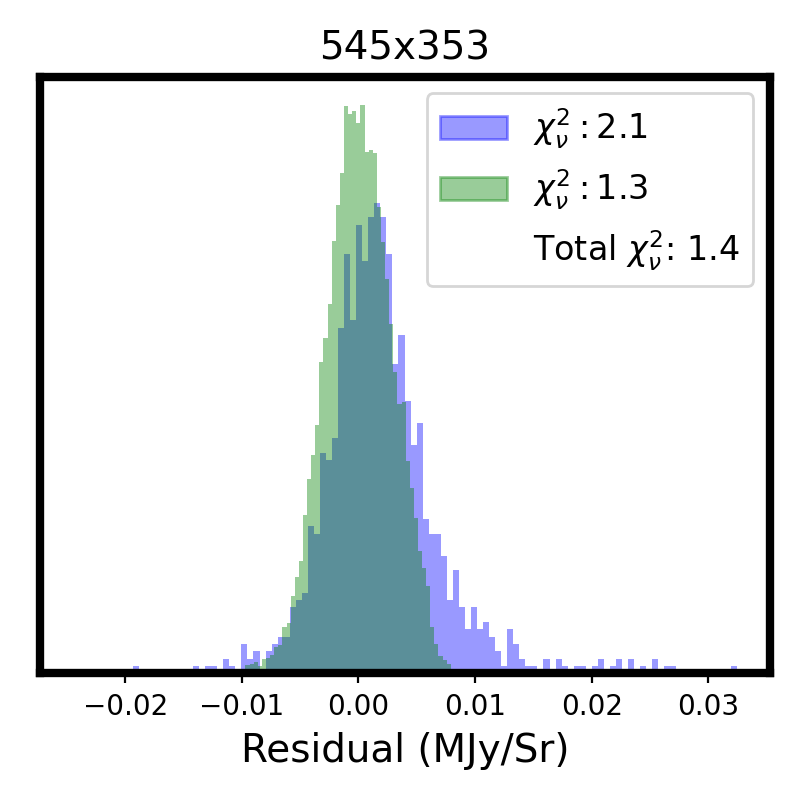}
\includegraphics[width = 0.24\textwidth]{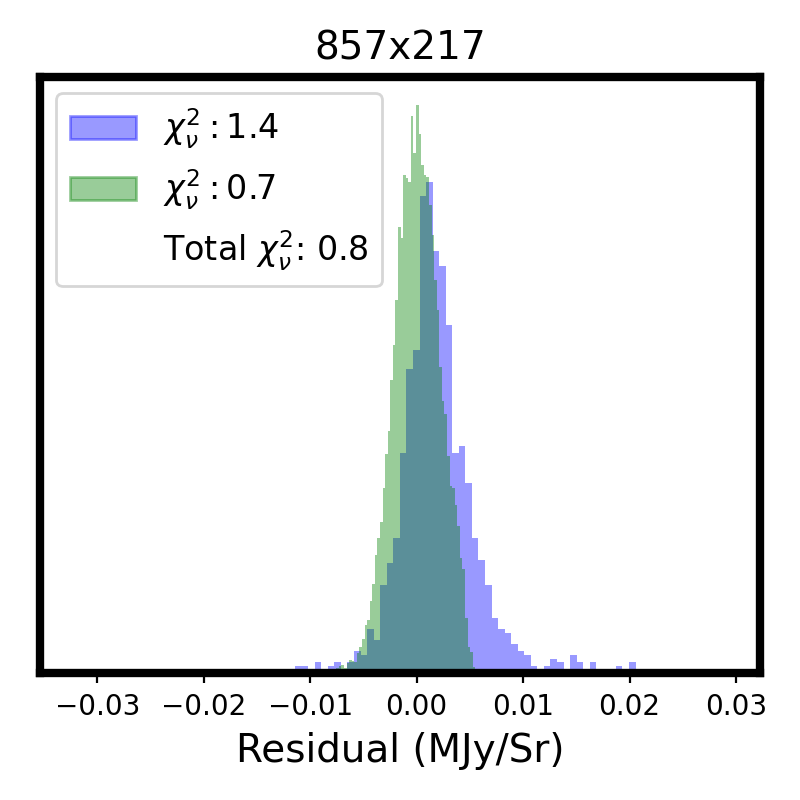}
\includegraphics[width = 0.24\textwidth]{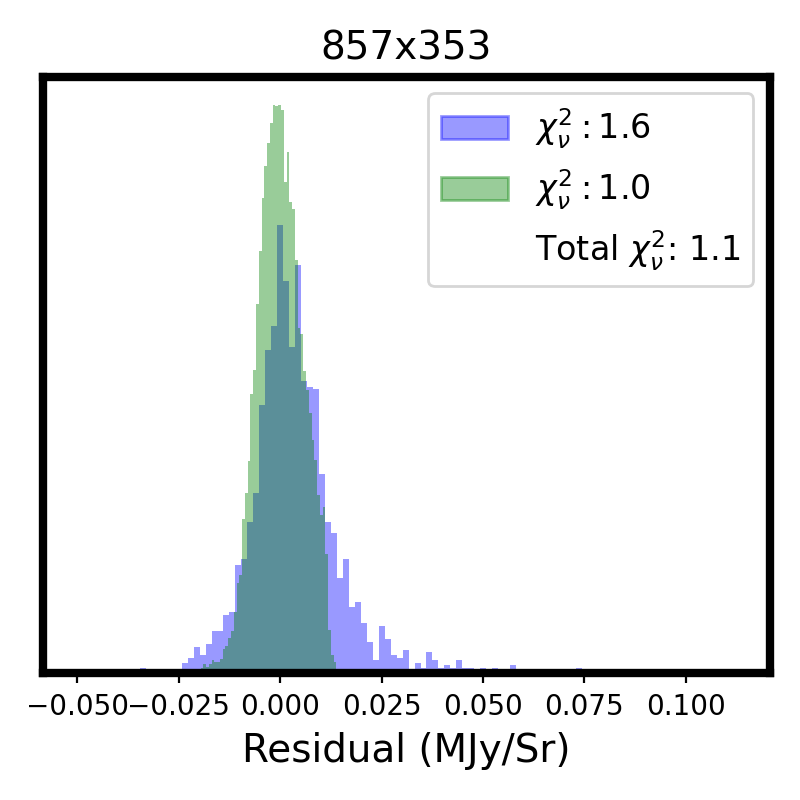}
\caption{Fitted populations over four different frequency pairings with the cleanest 40\% sky fraction. The top four panels show each data points in gray with errors in the abscissa and ordinate. Additionally, the data points in the panel are color-coded by $\pi_i$, representing the confidence that point belongs to each of the dust populations. The dashed line shows the population's mean value. The bottom four panels show the residual with respect to each population's mean. The y-axis shows the probability density, with a reduced $\chi^2$ computed for each population. These values can be compared to a single population ($K = 1$) fit which has values $\chi^2_\nu = [0.7, 1.3, 0.8, 1.0]$ in order from left to right.}
\label{fig:planck_scatterplot}
\end{figure*}

\begin{figure*}
\centering
\includegraphics[width=0.4\textwidth]{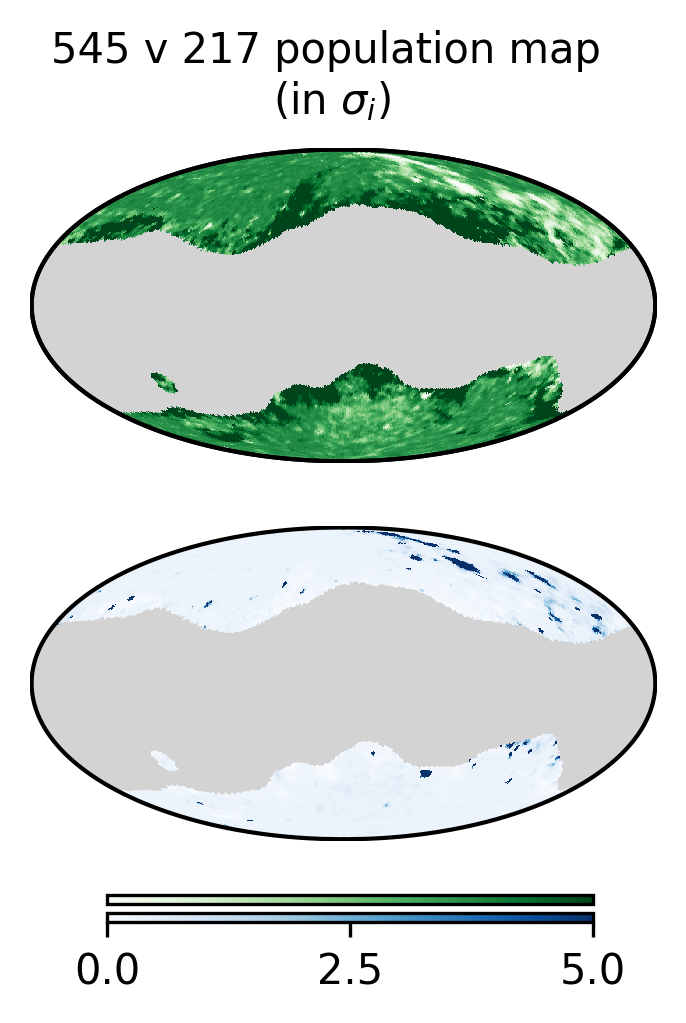}
\includegraphics[width=0.4\textwidth]{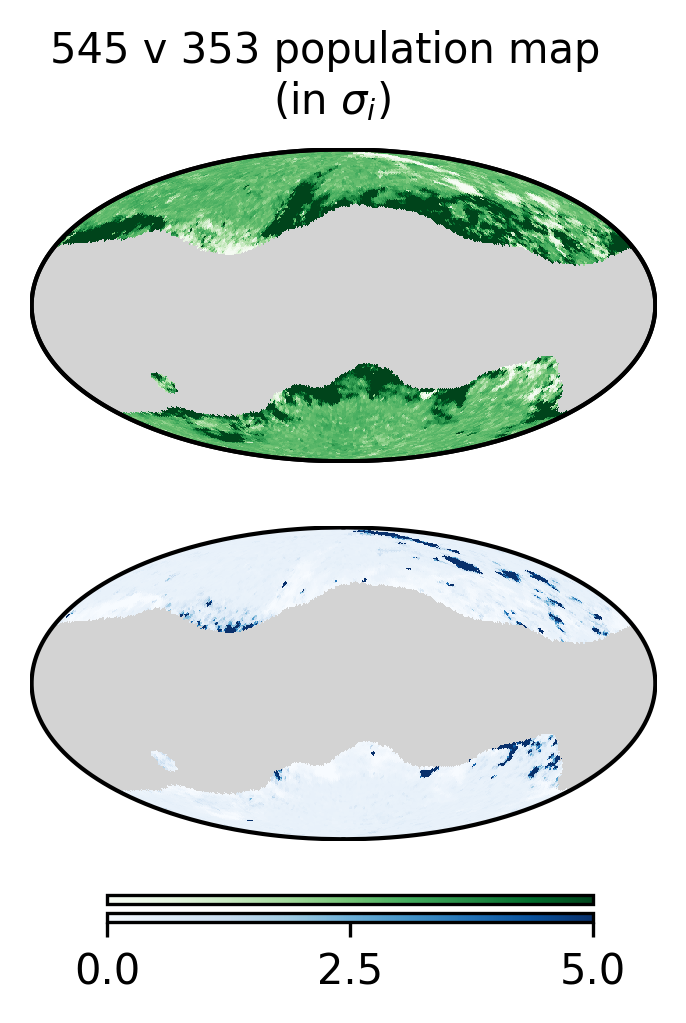} \\
\includegraphics[width =0.4\textwidth]{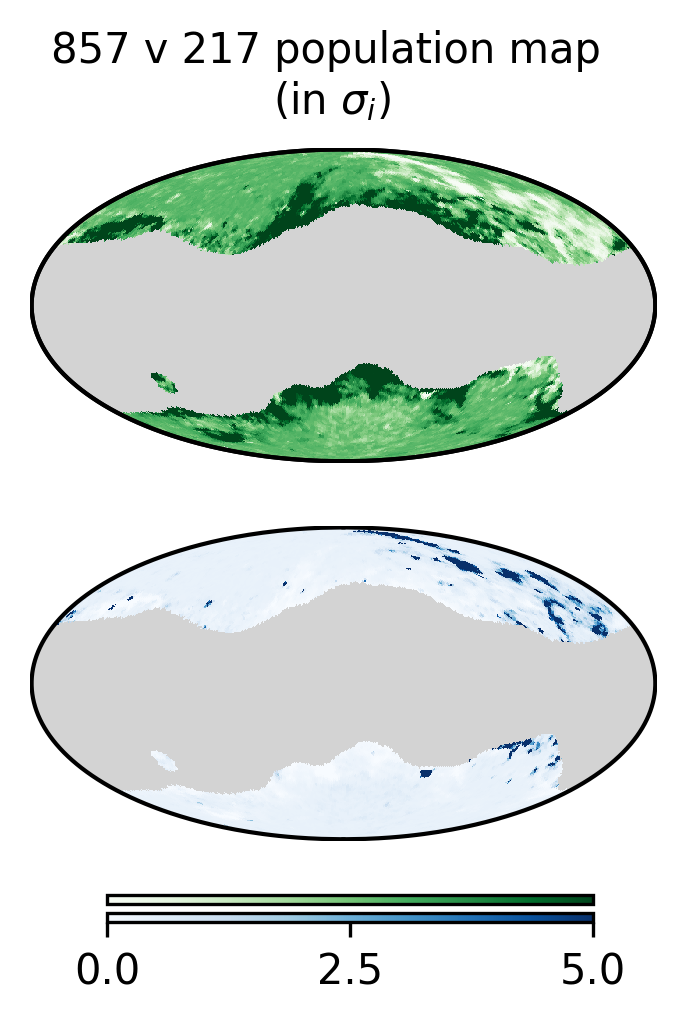}
\includegraphics[width =0.4\textwidth]{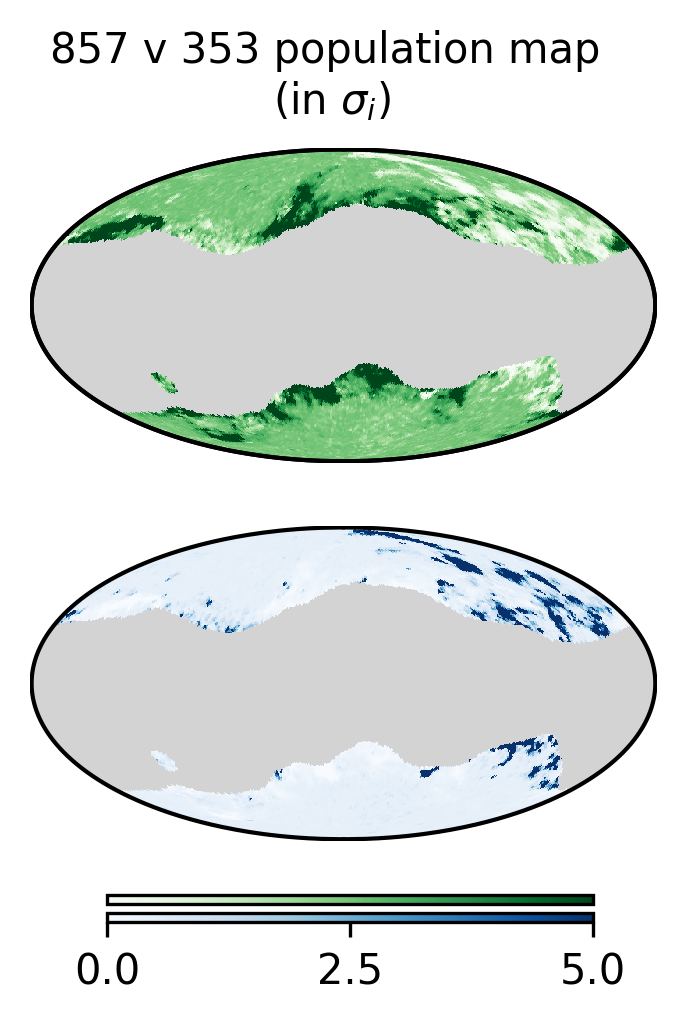}
\caption{Fitted dust populations over four different frequency pairings over the cleanest 40\% sky fraction. The likelihood has no inherent spatial information; nevertheless, in all scenarios, we identify two distinct dust populations with spatial coherence. The top (green) distribution is the dominant Galactic dust emission, with an increasing amount of statistical confidence the closer we are to the Galactic plane. The bottom (blue) population contains a different frequency scaling and consists of isolated dust clouds. } 
\label{fig:planck_double_mask}
\end{figure*}

We first conducted this analysis masking the Galactic plane \footnote{\lstinline{HFI_Mask_GalPlane-apo0_2048_R2.00.fits}}, keeping the cleanest 40\% of sky. We used the full likelihood as described in Equation \ref{eq:Kparamdust} with K=2 populations.
Physically, this model corresponds to two dust populations, with a mean SED frequency scaling $\theta_k$, and with some inherent variability inside each population parameterized by $V_k$. 
The role of modeling $V_k$ will be discussed in a future Section \ref{sec:dust_novar}. This model allows for miscalibration between frequency pairs and corrects for this by fitting a monopole calibration parameter.

While there is a preference for maximally uninformative priors, mixture models are highly multi-modal and susceptible to being stuck at local extrema \citep{chen_EM_convergence, wu2017convergence}. Therefore, in this analysis, we chose largely uninformative priors around realistic values, while not extending infinite support for all values. 

The slope was parameterized in angle from the $x$-axis and flat between $[0, \pi/2]$ to not prefer any slope aside from a positive relationship between dust emission. The monopole correction parameter has a truncated Normal distribution $b \sim \mathcal{N}(0, 1, \text{low} = -2, \text{high} = 2)$ that is practically flat across all relevant values.

We initially used an InvGamma($\epsilon$, $\epsilon$) prior for the variance due to its favorable properties, including (1) being a conjugate prior for a Gaussian, (2) support for infinite positive variance models, and (3) approximate flatness for small $\epsilon$ values. However, we discovered that this prior may lead to pathological fits. Sometimes, the fitted variance is driven by a few outliers and may be many times larger than the variance of the data itself. 
To address this, we instead selected the prior on the variance to a uniform distribution ranging from 0 to $2 \times$ the variance of the underlying data.

\smallskip 
\begin{table}
\captionsetup{size=footnotesize}
\caption{Fitted parameters for different pairs of analysis over the cleanest 40\% sky fraction and at $N_{\mathrm{side}} = 64$. Slopes are presented as $m = \tan(\theta)$, and the variance $V_k$ is the variance of the residuals. Note that the variance and slopes are in different units and cannot be directly compared. Instead, these should be compared to the noise levels of Table \ref{table:unit_conv} to see the increase in variance relative to the inherent noise level. The expected slopes between frequency pairs obtained with \Planck all-sky average values ($\beta_d, T_d$ = 1.53,19.6\,K \citep{Planck_2018_XI}) are [0.356, 0.102, 0.022, 0.287, 0.062] in descending order in the table. 
}
\label{table:fitted_LR40}
\centering
\begin{tabular*}{\columnwidth}{@{\extracolsep{\fill}}rcccccc}
\toprule 
  Freq. Pair & \multicolumn{2}{c}{Slope} & \multicolumn{2}{c}{Std. $(MJy/sr)$} & \multicolumn{2}{c}{Global frac} \\
 & $m_0$ & $m_1$ & $10^{3} \cdot \sqrt{V_0}$  & $10^{3} \cdot \sqrt{V_1} $ &  $p_0$ & $p_1$ \\ \midrule
 857$\times$545 & 0.328 & 0.362 & 10.3 & 17.3 & 0.67 & 0.33 \\
 857$\times$353 & 0.093 & 0.110 & 5.4 & 10.2 & 0.78 & 0.22  \\
 857$\times$217 & 0.021 & 0.027 & 1.8 & 3.2 & 0.84 & 0.16 \\ 
 545$\times$353 &  0.284 & 0.310 & 1.6&  3.2  & 0.85 & 0.15 \\
 545$\times$217 & 0.064 & 0.074 & 0.8 & 2.2 & 0.90 & 0.10  \\ 
 \bottomrule
\end{tabular*}
\end{table}

\smallskip

Figure \ref{fig:planck_scatterplot} shows the results of the fits to the data; Table \ref{table:fitted_LR40} summarizes the fitted parameters. 
The fitted slopes are consistent with our general expectation from \Planck all-sky average values $(\beta_d, T_d) \sim (1.53, 19.6\,\mathrm{K})$. 
We observe an alternate dust population across $\sim$15-20\% of the analyzed 40\% sky fraction. 
 
The fractional separation between the population's SED, $(m_1 - m_0)/m_0$, is more pronounced when the frequencies have a larger difference between them. 
This is consistent with a dust frequency scaling behavior because if the observed dust is composed of several populations, it becomes advantageous to make comparisons with a lower frequency map.
This is due to the longer lever arm it provides, facilitating the differentiation of various SEDs. 
However, there is a natural trade-off, because lower-frequency maps have lower signal-to-noise. At 857$\times$143\,GHz, the identification of multiple dust populations is not observed because the uncertainty for 143\,GHz is larger than the separation of the two populations. For this reason, the analysis with 143\,GHz or 100\,GHz is not included in Table \ref{table:fitted_LR40}. 

Every data point has a probability $\pi_{i,k}$ associated with it that describes the probability the point belongs to the $k^{\text{th}}$ population.
The bottom panels of Figure \ref{fig:planck_scatterplot} show the residuals between pixels and their assigned population, with reduced-$\chi^2_\nu$ values in the respective legends. 
These values, calculated for each model, are based on the most probable dust population assignments $\pi_{i,k}$. We leave the discussion of goodness-of-fit to Section \ref{sec:evidence}.

Because each point is treated as a Bernoulli process, each point can be converted to a statistical significance, 
\begin{equation} 
\sigma_{i,k} = \frac{\pi_{i,k}}{\sqrt{\pi_{i,k} (1 - \pi_{i,k}) }}.
\label{eq:sigma_i}
\end{equation} 
Then, we can transform these pixels back into the spatial domain and plot them in Figure \ref{fig:planck_double_mask}. 
Recall that no spatial information is built into the likelihood in Equation \ref{eq:Kparamdust}. Nevertheless, each dust population that is fitted shows a coherent spatial structure.  We observe a primary dust population in green, with high confidence, originating from dusty regions near the Galactic plane, and we identify isolated blue dust clouds that indicate the presence of a distinct alternate population.

\subsection{Comparison to known sources}
\label{sec:source_comparison}

We can compare our identified dust population to a dust extinction map. One such map is the Draine-Li 2007 dust extinction map\footnote{\lstinline{COM_ComMap-Dust-DL07-AvMaps2048_R2.00.fits}}. A visual comparison between this map and our most sensitive pair, 857$\times$353\,GHz, only shows a weak correlation. 
While some of the denser dust regions show up in the dust population mask, our analysis primarily identifies clouds on the eastward side of the Galactic plane. 
A simple Pearson correlation coefficient between a dust extinction map and our own analysis, over the same 40\% sky fraction, puts a correlation of $\rho = 0.32$. Clearly, dust density itself cannot be a good explanation for the observed population. 

SED variations are expected to naturally occur from gradients in the stellar radiation field with Galactic radius and height, leading to a heating of dust as we move towards the Galactic poles. However, this mechanism is expected to produce a symmetric pattern above and below the narrow Galactic disc.
Our observed population exhibits high asymmetry, which cannot be accounted for by this effect alone. 

Polarized radio loops are the largest structures in the sky and cause extended filamentary structures extending from the Galactic Plane \citep{Vidal_2015}. They are believed to be caused by supernovae remnants locally, and while brightest at radio frequencies from synchrotron radiation, they are also visible in the mm and sub-mm \citep{Liu_2014}. 
Indeed, the largest loops: I, II, III\footnote{Figure 2 of \citep{Vidal_2015}} are clearly missing in the identified dust population. However, smaller more localized bubbles may be responsible for some of the identified structures \citep{juan_superbubble}. A more systematic and careful comparison is warranted but beyond the scope of this paper. 

These large-scale SED variations are consistent with those found in \cite{Delouis_2021}, as shown in their Fig. 3, despite different analysis methodologies. 
Our identifications corroborate their conclusions, suggesting that these SED variations are more likely driven by large-scale Galactic physics rather than small-scale statistical variations in dust properties.

\section{Analysis over the \Spider Region}
\label{sec:spider_analysis}

\begin{figure*} 
\begin{minipage}[b]{0.38\textwidth}
\includegraphics[width = \linewidth]{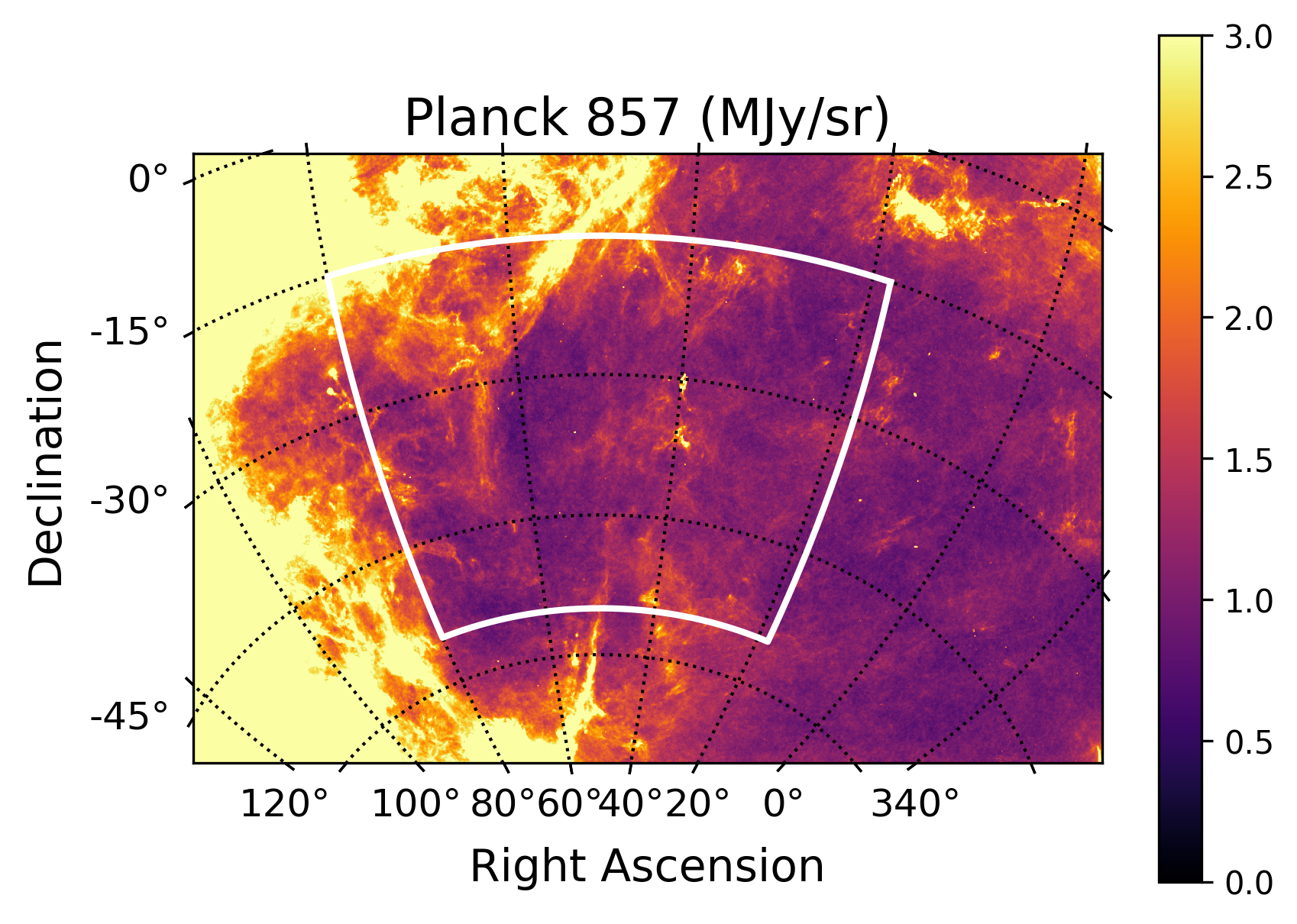} 
\vspace{-110pt}
\end{minipage}
\begin{minipage}{0.62\textwidth}
\includegraphics[width = \linewidth]{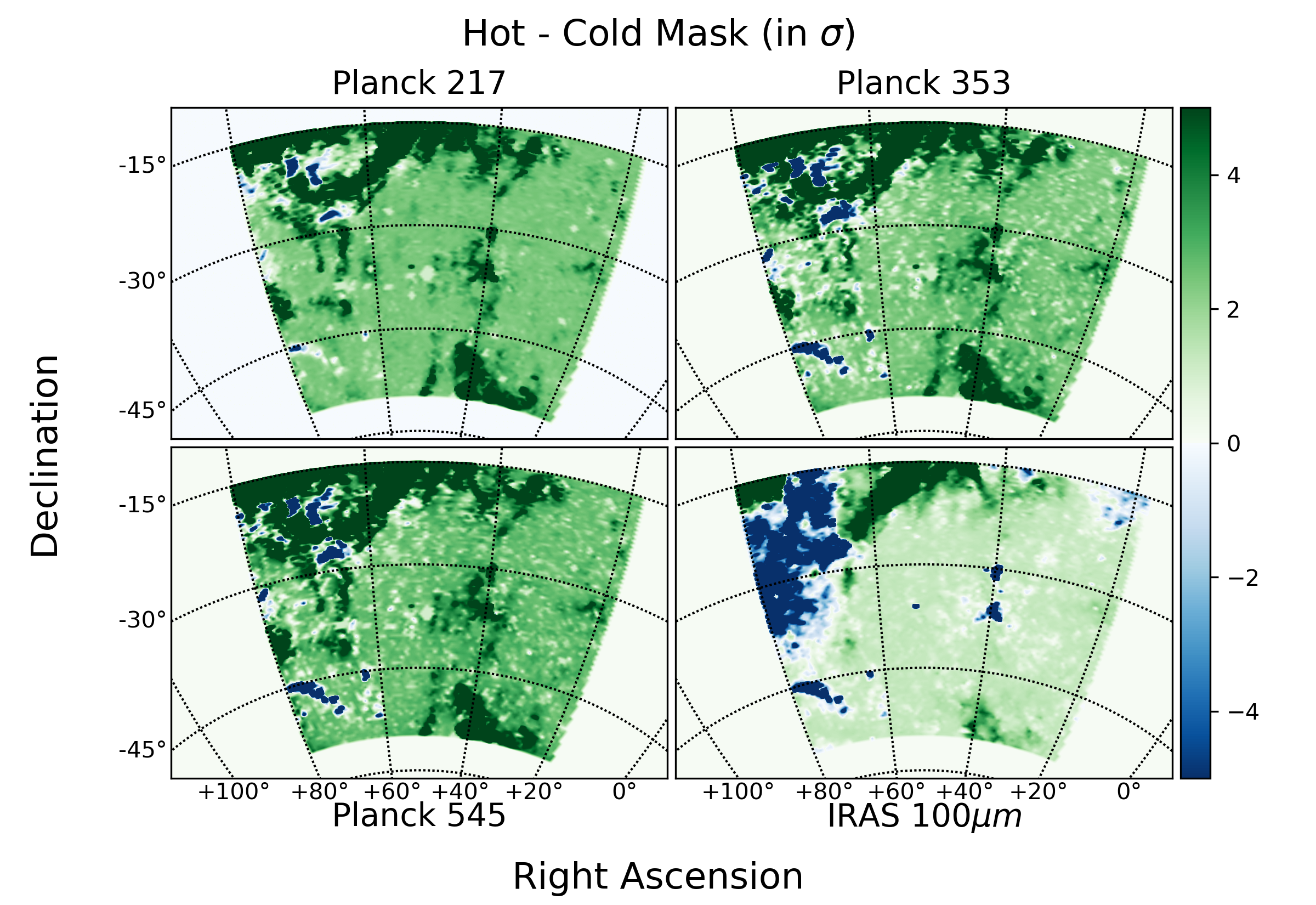}
\end{minipage}
\caption{Analysis of the dust populations over the \Spider observation region: a region in the Southern sky at Galactic latitudes greater than $|b|>20^\circ$. The leftmost panel shows \Planck's 857\,GHz map, which is a sensitive tracer of diffuse dust emission, over this region. The plots on the right show the identification of the dust populations, plotted as the difference $(\sigma_1 - \sigma_0)$ for conciseness. The four panel plot displays the analysis conducted with \Planck 857\,GHz and its corresponding frequency pair indicated above or below each subplot. With high confidence, we identify two populations of dust using any of the four input maps (\Planck 217,353, and 545\,GHz and \IRAS 100\,$\mu$m). We observe coherent dust clouds (in blue) that are interspersed between filamentary structures (in green). A visual comparison of the cloud structures with the dust intensity on the left shows little correlation between the two maps. } 
\label{fig:summary_hotcold_latlon_sigma}
\end{figure*}

The investigation of dust properties in high-Galactic latitudes holds considerable significance in the context of CMB studies. 
The conventional approach has focused on observing smaller sky areas, selected for reduced contamination from foreground sources \citep{spider_bmode, bicep_2018, Adachi_2022}.
We repeat this analysis over the \Spider region to identify if we can detect multi-populations of dust in this region: a high-Galactic latitude region $(|b| > 20^\circ)$ in the southern hemisphere comprised of $\sim$5\% of the full sky. This analysis is limited to dust \emph{intensity} rather than \emph{polarization}. Indeed, regions that are bright in polarization typically are the edges of dust clouds, where the magnetic field environments are more orderly than in the dense inner regions of clouds that are optically deep in intensity \citep{planckXIX2015}. However, the detection of a multi-population dust model in intensity would reveal that the foregrounds in this region are more complex than commonly modeled. 

The preprocessing pipeline is identical to that of Section \ref{sec:preprocess}, except the results presented here are at higher resolution $N_{\mathrm{side}} = 128$ and includes the ISSA reprocessing of the \IRAS 100\,$\mu$m data \footnote{\lstinline{ISSA_B4H0_healpix_ns1024.fits}} \citep{ISSA}.
The CMB is not subtracted from this map, as it's expected to be negligible at this frequency. Additionally, this map is smoothed to a common 10 arcmin resolution, which necessitates careful handling to avoid artifacts in the \Spider region.
ISSA maps provide coverage for 98\% of the sky, with the missing 2\% region coming close to the \Spider observation region. The missing region can be masked with a 3.75 deg apodization, chosen optimally to minimize ringing while avoiding masking the region of interest itself. 
Noise in the ISSA map is complex, although a uniform noise estimate has been tabulated as 0.07 \,MJy/sr$\pm$ 0.03\,MJy/sr \citep{ISSA, Miville_Deschenes_2005}. Out of caution, we adopt a noise level estimate that is 50\% larger than the quoted value, 0.105\,MJy/sr, uniformly across all pixels in the region. 

For this analysis, we use the same dust modeling as in Section \ref{sec:lr40_analysis}, where we fit two dust populations with a mean and variance for each.  
A summary of the results is shown in Figure \ref{fig:summary_hotcold_latlon_sigma} where four different frequency pairs are considered: \Planck 217, 353, and 545\,GHz and \IRAS $100\,\mu$m, all relative to \Planck 857\,GHz.

The fitted population fraction for the alternative model is $p_1 = 0.12, 0.15, 0.12, 0.34$, respectively, with $\sigma(p_1) = 0.01$ across all frequency pairs. 
In every case examined, we consistently observe coherent and filamentary structures across all pairs included in this analysis. 

The comparison of \Planck 857\,GHz with \Planck 545\,GHz is the most challenging due to the smallest frequency separation and, therefore, the most similar slope between the two populations. 
Indeed, in the analysis with 545\,GHz, the fit prefers a very similar slope between the two populations $m_0 \approx m_1$ with a substantially wider variance $V_1 \gg V_0$ which captures many of the outlier dust points.
Other analysis pairs instead distinguish themselves in $m_k$. 

For that reason, only a few points distinguish themselves in the pair with 545\,GHz. Conversely, \IRAS has the longest lever arm and also contains the most different spatial arrangement. 
Specifically, the shape of the dust cloud is different in the upper left region, and there is a disagreement in the population assignment in the central part of the region. 
ISSA has its own residual artifacts in its map, which differ from those of \Planck's. Therefore, identifying a common dust population strengthens the case for spatially separated dust populations in this region. 

We thus demonstrate that even the relatively clean \Spider region, known to be composed mainly of diffuse, non-turbulent dust \citep{planck_XLIV}, contains statistically detectable dust populations in intensity. 
While we expect the populations responsible for polarized mm-emission to differ from those that are bright in intensity, the existence of multiple dust populations in intensity suggests that dust is more complicated than commonly modeled.

\section{Robustness to Analysis Choices}
\label{sec:analysis_choices}

The likelihood function plays a central role in Bayesian parameter estimation. Therefore, it is crucial to check the appropriateness of the likelihood for the data at hand. In this section, we explore different likelihood choices to test the robustness of the conclusions. 

\subsection{Modeling dust without inherent scatter, $V_k = 0$}
\label{sec:dust_novar}

\begin{figure*} 
\begin{subfigure}[t]{0.19\textwidth}
\includegraphics[width = \textwidth]{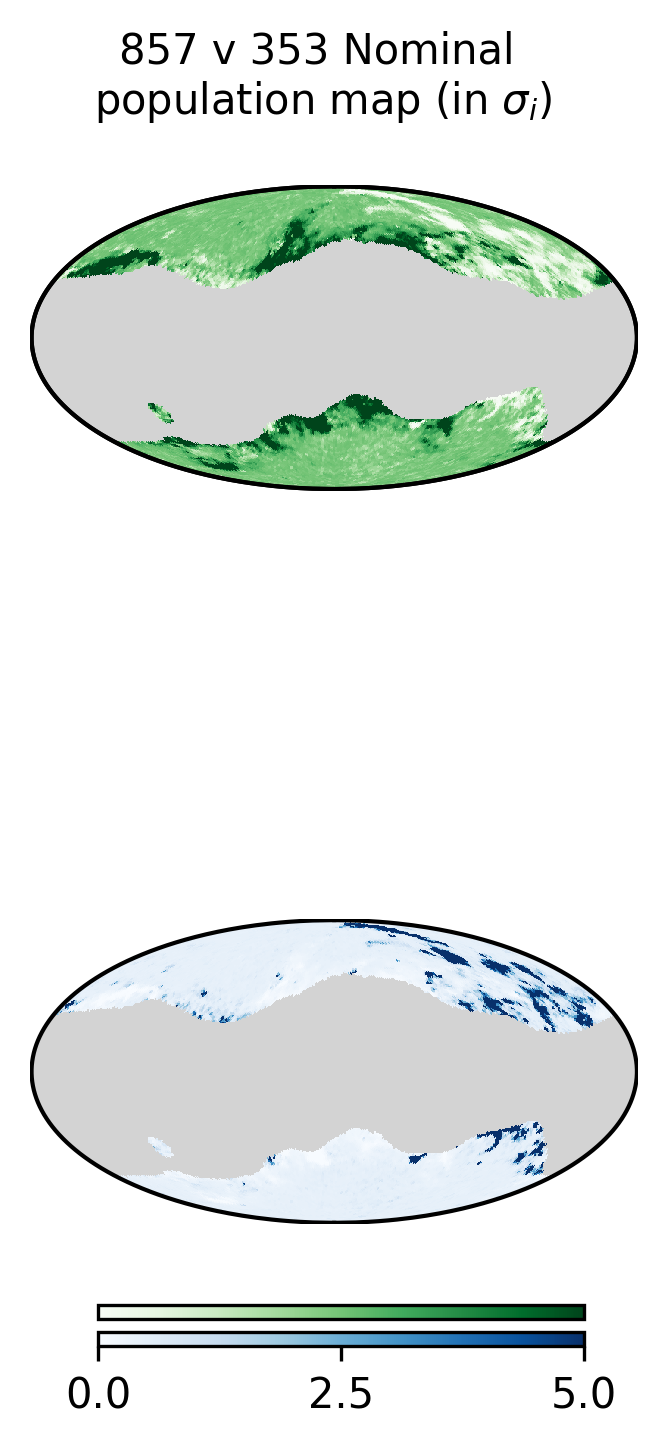}
\caption{}
\end{subfigure} 
\begin{subfigure}[t]{0.19\textwidth}
\includegraphics[width = \textwidth]{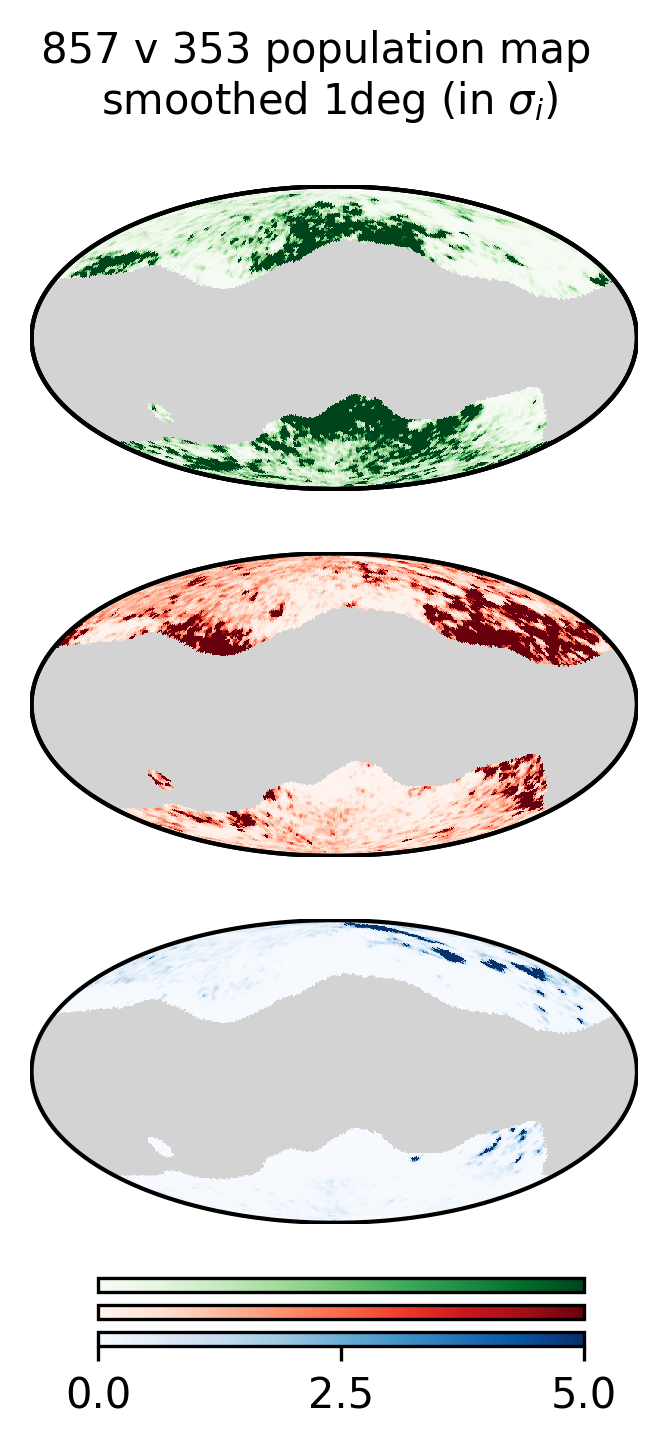}
\caption{}
\label{fig:analysis_choices_smoothed}
\end{subfigure}
\begin{subfigure}[t]{0.19\textwidth} 
\includegraphics[width = \textwidth]{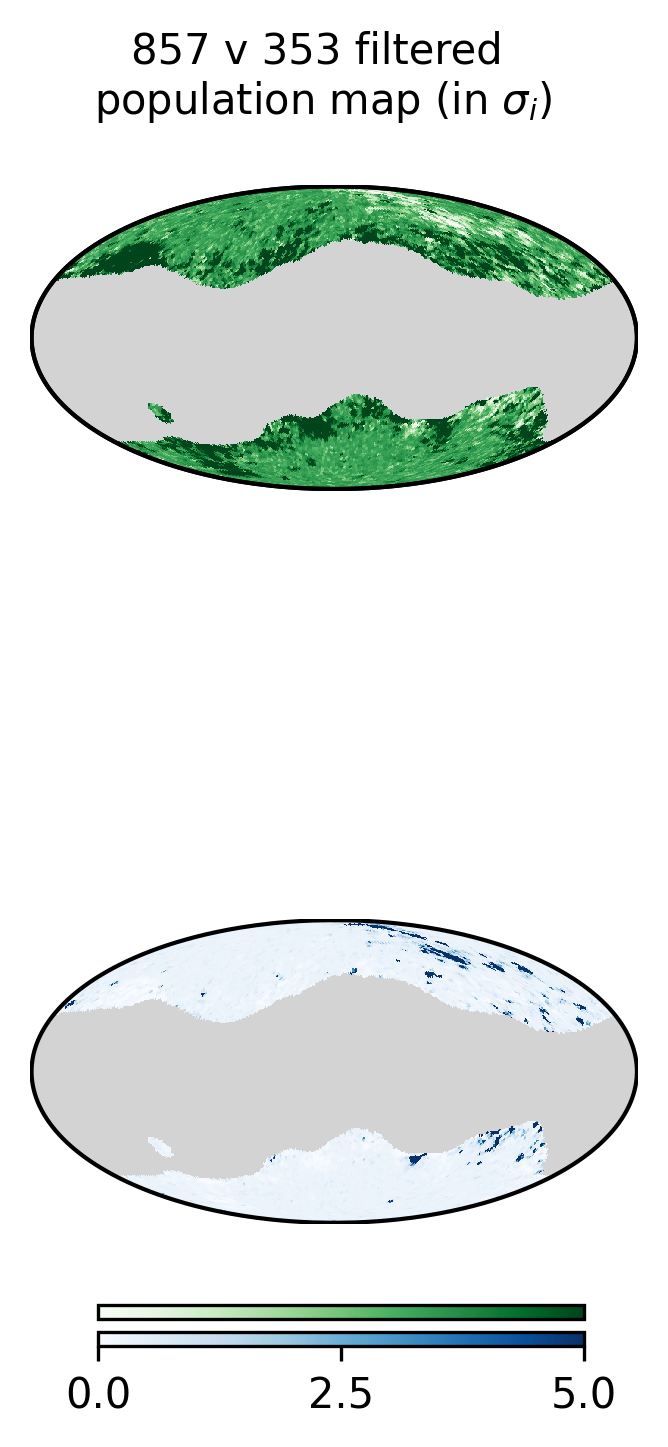} 
\caption{}
\label{fig:analysis_choices_filtered}
\end{subfigure}
\begin{subfigure}[t]{0.19\textwidth} 
\includegraphics[width = \textwidth]{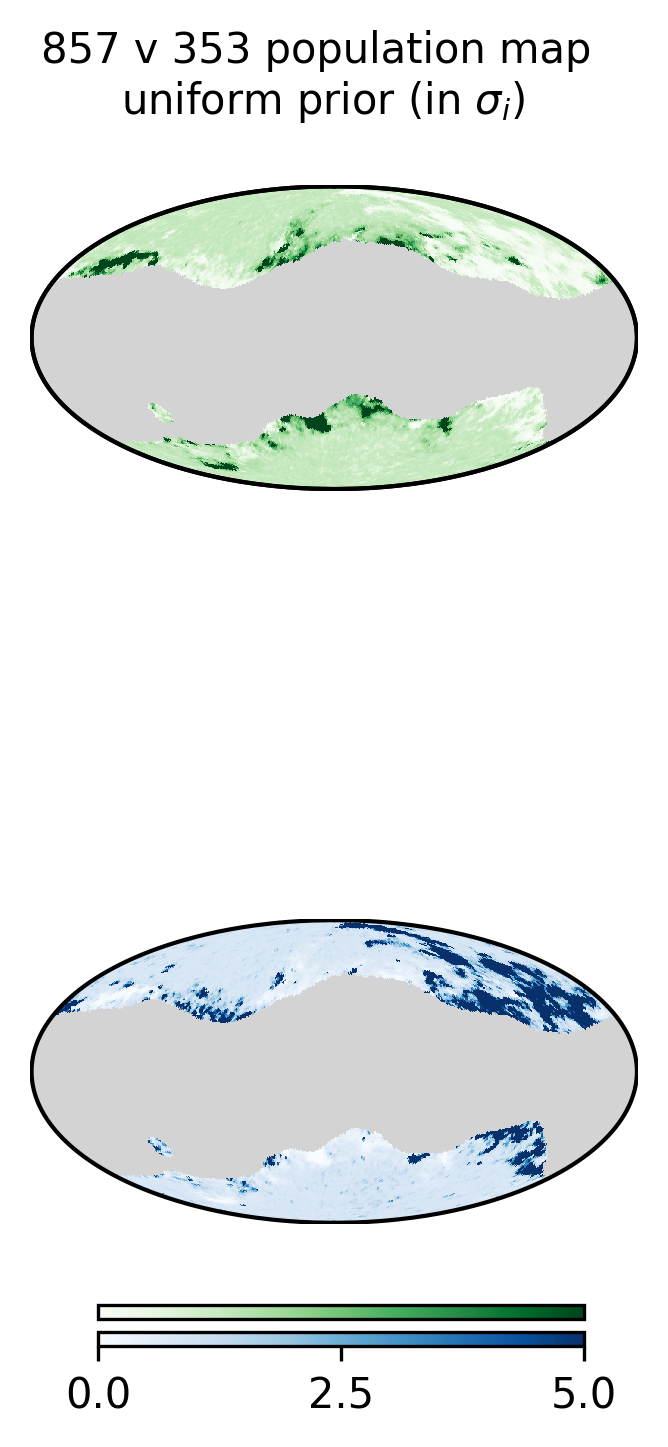} 
\caption{}
\label{fig:analysis_choices_noprior}
\end{subfigure}
\begin{subfigure}[t]{0.19\textwidth} 
\includegraphics[width = \textwidth]{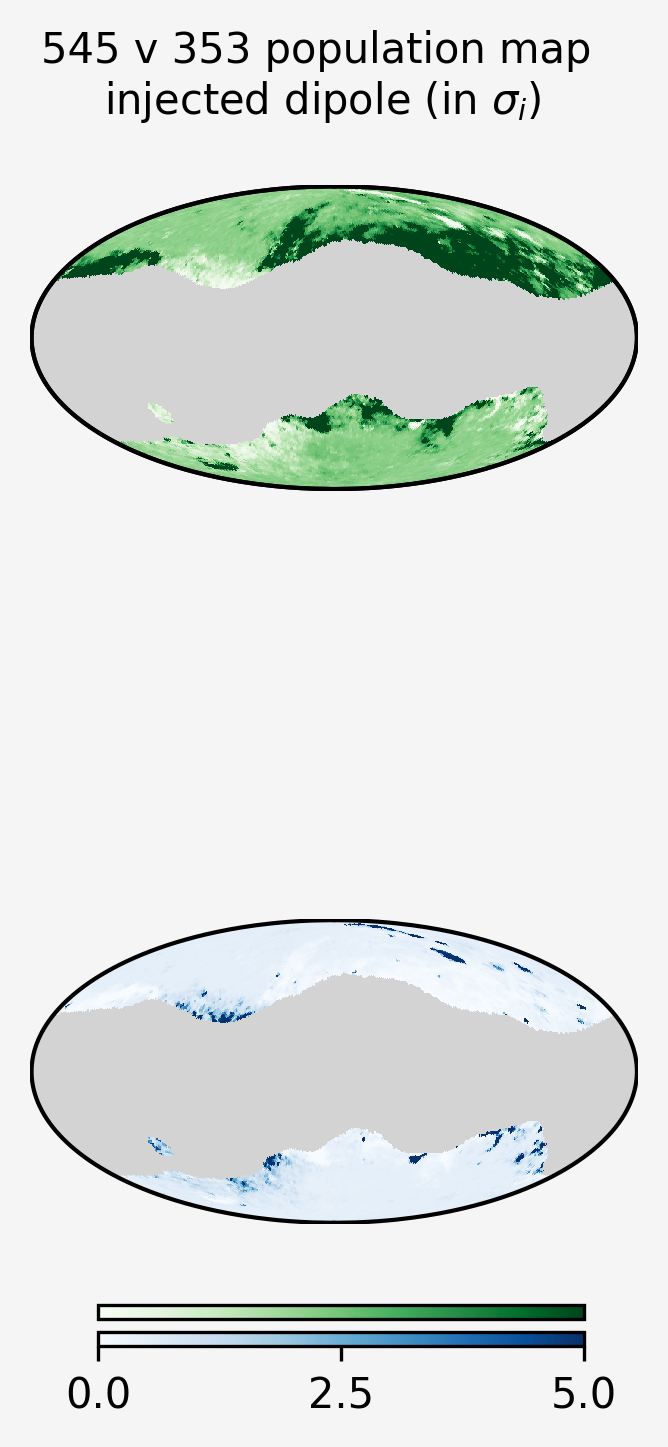}
\caption{}
\label{fig:analysis_choices_dipoleadded}
\end{subfigure}
\caption{Investigating alternate analysis choices and their effect on the identification of multiple dust populations. (a) Reference figure copied from Figure \ref{fig:planck_double_mask} for the basis of comparison. (b) Likelihood model of dust composed of discrete populations with no intrinsic variations, $V_k = 0$. A beam smoothing factor with a FWHM = 1deg is applied to the $\pi_i$ to effectively apply a spatial weighting factor. Coherent structures in this scheme suggest that spatial information may allow the separation of more components, however this extension was not pursued. (c) Filtered maps that remove large-scale artifacts in the map. The variance is included, $V_k$, while the monopole offset is not needed $(b_k = 0)$. We see a similar spatial arrangement of our two dust populations, indicating that large-scale residual effects, primarily the residual solar dipole, do not inadvertently influence our identification of two dust populations. (d) No hyperparameter prior $p_A$. In this configuration, $\pi_{i}$ is drawn from a uniform distribution from 0 to 1. (e) An analysis adding back a 10\% solar dipole signal to the 545\,GHz map, introducing a dipole residual between the pair of maps. A comparison to Figure \ref{fig:planck_double_mask} reveals similar populations, indicating this effect cannot be an artifact stemming from inherent dipole residuals. Note that this mask utilizes different inputs compared to its neighbors. }
\label{fig:analysis_choices}
\end{figure*}

In Section \ref{sec:lr40_analysis}, our likelihood model incorporates a term denoted as $V_k$, representing the variance specific to each population. This has a physical interpretation within the model: we are studying different dust populations, and these populations are generated through a process characterized by an average scaling value, $\theta_k$, with inherent variations about that mean value captured by $V_k$. 

In this section, our objective is to investigate whether the data can be accurately represented using only a single spatially constant, $\theta_k$, with a monopole offset. In this model, any observed variance is attributed solely to measurement noise. 

To account for the variation in data points, we find that the data prefers a fit with three slopes. 
When we look at our most significant dust pair, 857$\times$353\,GHz, we have three distinct dust populations, with fitted population fractions of \{0.47, 0.46, 0.07\} from the shallowest to the steepest slope. These proportions remain true for several different frequency pairs. 
 
In cases where there is insufficient statistical power to fit three dust populations and instead, two populations are fit, the likelihood analysis prefers to lump populations 0 and 1 together and still distinguish population 2 separately. This is evident when analyzing 545$\times$217\,GHz. 

In the likelihood analysis, each pixel is assigned a probability $\pi_i$ of belonging to one of K populations. 
The probability is determined by its distance to each population based on the likelihood ratio: $\pi_{i,k} = \mathcal{L}_i(k)/\sum_k \mathcal{L}_i$. 
The likelihood function, comprised of a sum of Gaussians, makes the likelihood ratio highly sensitive to the distance of each data point from its nearest population. 

 Within this framework, the impact of statistical noise inadvertently magnifies. Points exhibiting fluctuations near the population boundaries draw exponentially closer to one population over another, leading to $\pi_i$ assignments that are overly confident. 


We can make use of spatial arrangements of pixels to suppress the effects of noise. When multiple neighboring pixels agree on a population, then this consensus strengthens our confidence in the correctness of the assignment. Alternatively, if neighboring pixels conflict, we should suppress the assignment. 

A simple way to implement this \emph{a posteriori} is to apply a smoothing factor on the probability map $\pi$. We smooth the map with a symmetric Gaussian beam, and then calculate the $\sigma$-significance in the same way (Eq. \ref{eq:sigma_i}).

Figure \ref{fig:analysis_choices_smoothed} shows the results of our most sensitive pair, 857$ \times$353\,GHz, when smoothed with a 1-degree beam. These are sorted from top to bottom , representing SED scalings ranging from the shallowest (green), to steepest (blue). In this order, we observe dominant dust emission concentrated around the Galactic plane, followed by diffuse Galactic emission at the edges of the Galaxy closer to the poles. The last population is isolated cloud structures that deviate from the general dust population. 

This analysis reveals insights about the arrangement of dust data points in the scatterplot (Fig. \ref{fig:planck_scatterplot}). Notably, the stratification of pixels in the scatterplot space holds physical significance. Points with lesser slopes come from the brightest regions of the Galaxy, while points with greater slopes come from more diffuse regions of the galaxy. 

This analysis suggests that incorporating spatial information into the likelihood equation could potentially yield additional insights into dust populations.  
Such an extension would require us to encode the spatial relationship between pixels in the likelihood.  
Neighboring pixels may collectively indicate the presence of an alternate population, even if individual pixels lack sufficient statistical significance for identification. 
Such an analysis would be a substantial increase in modeling complexity. 
In intensity, we have high-frequency data with strong signal-to-noise ratio, allowing us to differentiate points without relying on spatial information. 
Instead, this extension could prove valuable in the study of polarized foregrounds. However, it falls outside the scope of this paper to introduce spatial modeling. 

\subsection{Removing dipole residuals with harmonic space filters}
\label{sec:filter}

There are several challenges with analyzing native \Planck maps. \Planck maps contain an irreducible zero-level offset. The monopole offset has physical sources, including cosmic infrared background and zodiacal emission \citep{planck_III_2018}. Typically this offset is adjusted based on tracers from other datasets, such as HI \citep{planck_III_2018}. 

Additionally, the \Planck maps contain residual artifacts from the Solar dipole: the motion of the Solar System relative to the rest frame of the CMB. The major effect of residual dipole occurs at $\ell = 1$. However, relativistic corrections will leak into smaller scales $\ell > 1$ \citep{planck_VIII_2015}.
Bracketing the effect of dipole residuals is especially important as the dominant population occurs in the region of the dipole maximum.


One possibility is to modify Equation \ref{eq:Kparamdust} to include a residual dipole term. This dipole would be aligned in the direction of the Solar dipole ($\ell, b$) = (264.02$^\circ$, 48.253$^\circ$), and the model would include a term to fit for a mismatch in dipole residuals between a pair of frequency maps. 
We have performed this exercise and recovered mismatch amplitudes of no more than 0.7\% when analyzing mm-channels (545, 857\,GHz) with the sub-mm channels (217,353\,GHz) and 2.3\% when analyzing the mm-pair 545 vs 857\,GHz. 
In all situations, we are able to reproduce the alternate dust population as demonstrated in Figure \ref{fig:planck_double_mask}. 
This potential solution would not capture any drifts in the dipole direction between the mm-channels vs the sub-mm channels which is expected to be a major source of uncertainty for the mm-channels. 

Instead, the approach in this section is to circumvent these issues by filtering the maps to eliminate large scale features and fit only dust anisotropies. 
We apply an aggressive window function that removes all power below $\ell \leq 8$, effectively eliminating all possible large-scale dipole residuals. To reduce the impact of undesired ringing, we choose a cosine apodization around $\ell = 20$ to smoothly taper the filter's response function, 

\begin{equation}
  W(\ell) =\begin{cases}0   \qquad & \text{ if } \ell \leq 8, \\
  0.5 \left(1 - \cos \left( \frac{ 2 \pi ( \ell - 8)}{50} \right) \right) \qquad  & \text{if } 8 <\ell \leq 33, \\
  1 & \text{if } \ell > 33.
  \end{cases}
\end{equation}
Additionally, we take precautions to prevent bright Galactic plane emissions from causing unwanted leakage into the rest of the map, which would inadvertently bias our analysis. For this reason, we apply the filter to 60\% of the sky, ensuring minimal ringing in our 40\% sky fraction, which is the specific region of interest for this analysis. 

With these preprocessing steps, no offsets are needed in Equation \ref{eq:Kparamdust}; therefore, we set $b_{k} = 0$ for all $K$ populations. We include $V_k$ in this version in order to compare it to our nominal case. 

We fit the filtered maps with $K = 2$ populations, resulting in two distinct slopes. Figure \ref{fig:analysis_choices_filtered} visualizes the results in the spatial domain. 
We see a spatial arrangement of pixels similar to the unfiltered analysis: 
A dominant diffuse population alongside an alternate population localized to specific clouds. This is consistent with the analysis in Section \ref{sec:lr40_analysis}; therefore, we believe that large-scale residual artifacts in \Planck cannot be the origin of these SED variations.

\subsection{Bracketing the effect of solar dipole residuals}
\label{sec:bracket}

A potential concern with Section \ref{sec:filter} is that the removal of large scales depends on both the chosen mask and filter parameters. 
\Planck 545 and 857\,GHz have approximately 5\% calibration uncertainty, resulting in a residual CMB dipole amplitude uncertainty of up to 0.05$\times 3361 \mu\mathrm{K_{cmb}} = 168\mu \mathrm{K_{cmb}}$ \citep{planck_III_2018}. 
In this section, we add a dipole pattern aligned with the true solar dipole ($\ell, b$) = (264.02$^\circ$, 48.253$^\circ$) with an amplitude 10\% of the peak value, representing a residual twice as large as our expectations. This dipole residual is converted to flux-density units (Table \ref{table:unit_conv}), and added to our mm-map to mimic the presence of dipole residuals. 
This is propagated through our analysis to ascertain whether residual dipoles could potentially explain the observed SED variation. 
If these populations are no longer discernible, this suggests that SED variations may be explainable by residual solar dipole contamination. 
 
Across all pairs of maps, we recover similar identification of dust populations. Figure \ref{fig:analysis_choices_dipoleadded} shows the analysis with 545\,GHz, which has an substantially larger effect than the same injection into 857\,GHz due to the shape of the blackbody curve at the CMB temperature. 

A comparison of this population with the base analysis, Figure \ref{fig:planck_double_mask}, shows similar general structure of identified dust clouds. 
The dust clouds that are near the peak of the solar dipole direction have diminished in significance. The cloud structure from the Northern pole has deviations larger than 10\% of the solar dipole signal and remains well identified. 
Additionally, the clouds in the Southeastern quadrant are unaffected by a dipole residual and remains well identified. The injection of this solar dipole at the dipole minimum, in the Southwestern quadrant, creates an artifact that is otherwise picked up. 
This test, representing an exaggerated effect from solar dipole residuals, shows that our two populations identified remains robust and cannot be explained by solar residual artifacts. 

\subsection{Removing hyperparameter $p_K$}
\label{sec:noprior}

In Section \ref{sec:sims}, we demonstrated the necessity of the hyperparameter $p_A$ for $q_{i} \sim \text{BetaBinom}(n = 1, \alpha = p_A, \beta = 1 - p_A)$ to ensure an unbiased recovery of our simulated inputs. 
However, in this section, we intentionally omit the hyperparameter to observe the impact on this analysis. In this configuration, $q_i \sim \text{BetaBinom}(n = 1,\alpha = 1, \beta = 1)$. Effectively, we first draw $\pi_{i,A}$ from a Uniform(0,1) distribution. Then, $q_i$ is drawn from a Bernoulli distribution with $p = \pi_{i,A}$. 

The outcomes of this configuration, as depicted in Fig. \ref{fig:analysis_choices_noprior}, reveal a similar spatial arrangement of pixels. 
The similarity arises because data points that show substantial separation between our two populations possess sufficient statistical power to remain largely unaffected by the absence of the prior. 
Indeed, a simple calculation of the Pearson correlation coefficient between the nominal case and this case is 0.88. The scatter plot of $p_i$ between both scenarios demonstrates a monotonic but non-linear relationship. Therefore, despite a large difference in analysis choice, we maintain a relatively high correlation coefficient. 


However, points that fall statistically in-between populations are instead heavily influenced by the choice of prior. In this case, points are assigned evenly between two populations, causing the global slopes to converge and leading to an over-identification of the less numerous blue population. 
This behavior is evident in the simulations (Fig. \ref{fig:nuts_example}) where there is a substantial bias in the globally fitted slope and a greater assignment of the alternate population. 
On the data, the alternate population $\geq 3\sigma$  expands substantially from 6\% to 15\%. 

While a uniform distribution is often thought to be uninformative, it's important to note that its mean is fixed at 0.5. 
This fixed mean influences $\pi_{i,A}$ to tend towards larger values compared to a scenario where the hyperparameter $p_A$ is allowed to adapt the mean based on what is preferred by the data. Consequently, using a uniform prior leads to a \emph{less conservative} identification of an alternate population. For this reason, and because it fails to recover inputs on simulations, we opt not to use a uniform prior on $\pi_{i,A}$ for this analysis.

\section{Bayesian Evidence for the Models}
\label{sec:evidence}
The previous sections have demonstrated there is a lot of choice in the models we can fit to our data. 

A simple metric is $\chi^2_\nu$ to assess goodness-of-fit to each model. In Figure \ref{fig:planck_scatterplot}, the legends display the residuals relative to each model along with the $\chi^2_\nu$ for each individual population. The total $\chi_\nu^2$ combines both models to assess the global model.
The blue populations have $\chi^2_\nu > 1$, which may suggest that the variance $V_k$ is underestimated for this alternate dust population.
However, there are two primary effects leading to a larger $\chi^2_\nu$. (1) The residuals are not centered around 0. This occurs because the two populations are truncated at $\pi_{i,k} = 0.5$, leading to a shift in the residual distribution.
(2) Additionally, there are some large outliers that are not captured by this simple two-population model. In the blue population, there are outliers with residuals that exceed $5\sigma$ from the population, occurring more frequently at approximately the $\sim 3 \sigma$ level. 

A one-population model has competitive goodness-of-fits as highlighted in the caption of Figure \ref{fig:planck_scatterplot}. However, this one-population model requires a larger $V_k$ to accommodate the width of the observed data.
Indeed, in the model, $V_k$ can accommodate any model with a good $\chi_\nu^2$, at the expense of a low likelihood.

Rather than goodness-of-fit, we ask which of the models has the highest marginal likelihood, or the highest Bayesian evidence to select the correct model to fit. A Bayesian evidence calculation naturally  balances between model complexity and goodness-of-fit: more components will always return a higher likelihood value, but the larger prior volume can diminish the evidence if the increase in parameter space does not adequately represent the data.

Calculation of Bayesian evidence is computationally expensive and requires a sum (of Eq. \ref{eq:Kparamdust}) over $K^N$ states of $q_i$. This is simply intractable for our pixelization and sky area. Instead, we can perform an analytical trick to simplify the calculation of the marginal likelihood. Following the method of \citep{hogg_likelihood}, each $q_i$ has K possible states with probability $\{p_1,\cdots p_k \}$ so the likelihood, marginalized over $q_i$, can instead be written as a sum of states,
\begin{equation}
  \mathcal{L}( \theta_k, b_k, V_k, p_k) = \prod_i  \sum_k \left[ \frac{p_k}{\sqrt{2 \pi (\Sigma_{k,i} + V_k) }} \exp \left( - \frac{\Delta_{k,i}^2}{2 (\Sigma_{k,i} + V_k)} \right) \right], \\  
  \label{eq:Kparamdust_marginalized}
\end{equation}
where now the likelihood only has $4K - 1$ parameters. 

We use the nested sampler \emph{dynesty} \citep{dynesty_citation, dynesty_software} to perform this calculation with results in Table \ref{table:bayesian_evidence}. The statistical and sampling uncertainty for all the quoted values is subdominant to the significant figures in the table. 

The preferred model, for the majority of the pairs, is a two-population ($K = 2$) model that include an additional variance term $V_k$. For this reason, the quoted results in Section \ref{sec:lr40_analysis} are those of this analysis. However, as we go to lower frequency pairs, the signal-to-noise is lower on dust and simpler models are adequate to explain the variations in the data. For $545 \times 217$, a multi-population dust model with no additional variance $(V_k = 0)$ has the highest evidence. 

One could combine all datasets \Planck 100-857\,GHz and \IRAS 100\,$\mu$m and construct a multi-variate likelihood to fit them simultaneously. Such an approach would not significantly increase the number of fitted parameters, because the parameter $q_i$ is associated with each spatial pixel and would not scale with the number of inputs. However, since our pairwise analysis shows a clear distinction between the two populations, we have chosen not to pursue this extension at this time.

\begin{table}
%
\captionsetup{size=footnotesize}
\caption{Calculation of marginal likelihood over 40\% of the sky with $N_{\mathrm{side}}=64$ using two different models: inc $V_k$ as described in Section \ref{sec:lr40_analysis} and $V_k = 0$ as described in Section \ref{sec:dust_novar}. All values of marginal likelihood are in $\log_{10}$ units. The highest evidence values are bolded. The highlighted blue background provides a qualitative comparison of the log marginal likelihood values in relation to the highest value in the row. }
\begin{tabular*}{\columnwidth}{r@{\hskip 0.2cm}c@{\hskip 0.1cm}c@{\hskip 0.3cm}c@{\hskip 0.1cm}c@{\hskip 0.1cm}c}
\toprule 
& \multicolumn{2}{c}{Inc. $V_k$} & \multicolumn{3}{c}{$V_k = 0$} \\
Frequency pair & $K = 1$ & $K=2$ & $K=1$ & $K=2$& $K=3$\\ \midrule
857$\times$545\,GHz & 
\cellcolor{blue!23}$1.3\cdot10^4$&\cellcolor{blue!25} $\mathbf{ 1.4\cdot10^4}$ &\cellcolor{blue!0} $-3.7\cdot10^5$&\cellcolor{blue!0} $-1.3\cdot10^5$&\cellcolor{blue!0} $-7.5\cdot10^4$\\
857$\times$353\,GHz & \cellcolor{blue!19} $4.0\cdot 10^3$&\cellcolor{blue!25}$\mathbf{5.4\cdot 10^3}$&\cellcolor{blue!0}$-5.0\cdot10^4$&\cellcolor{blue!0}$-1.4\cdot10^4$&\cellcolor{blue!0} $-5.6\cdot 10^3$\\
857$\times$217\,GHz & 
\cellcolor{blue!11} $7.5\cdot10^2$&\cellcolor{blue!25} $\mathbf{1.7\cdot10^3}$&\cellcolor{blue!0}$-9.9\cdot10^3$&\cellcolor{blue!0}$-1.6\cdot10^3$&\cellcolor{blue!0}$-6.4\cdot10^2$\\
545$\times$353\,GHz & 
\cellcolor{blue!21} $1.1\cdot10^4$&\cellcolor{blue!25}$\mathbf{ 1.3\cdot10^4}$&\cellcolor{blue!14} $7.4\cdot10^3$&\cellcolor{blue!23} $1.2\cdot10^4$&\cellcolor{blue!23} $1.2\cdot10^4$\\
545$\times$217\,GHz &
\cellcolor{blue!22} $4.5\cdot10^3$&\cellcolor{blue!23} $4.6\cdot10^3$&\cellcolor{blue!24} $4.7\cdot10^3$&\cellcolor{blue!24} $4.9\cdot10^3$&\cellcolor{blue!25} $\mathbf{5.0\cdot10^3}$\\
\bottomrule
\end{tabular*}
\label{table:bayesian_evidence}
\end{table}

\section{Inferred Galactic Dust Properties}
\label{sec:inferred_dust}
\begin{figure*}

\includegraphics[width = 0.5\textwidth]{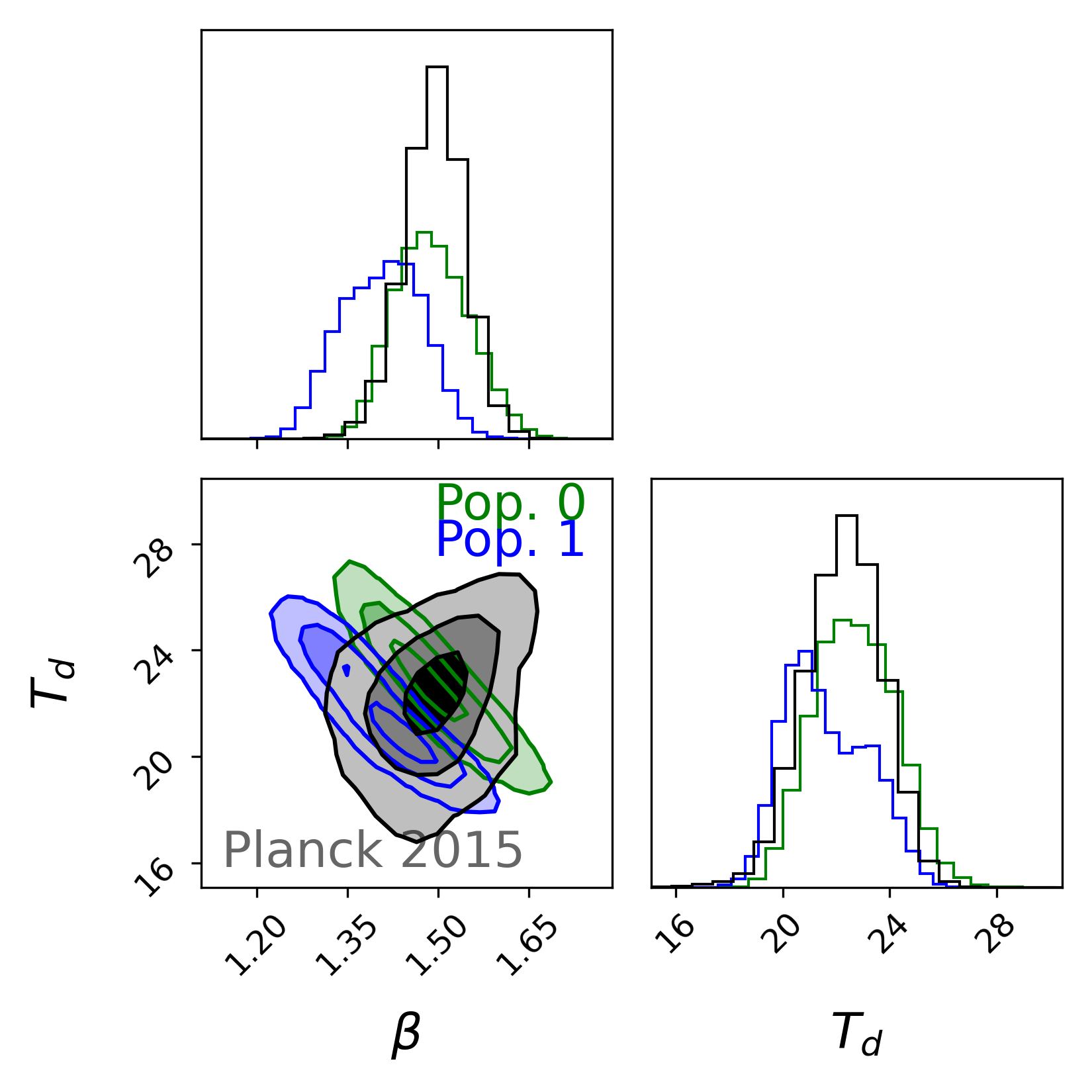}
\includegraphics[width = 0.5\textwidth]{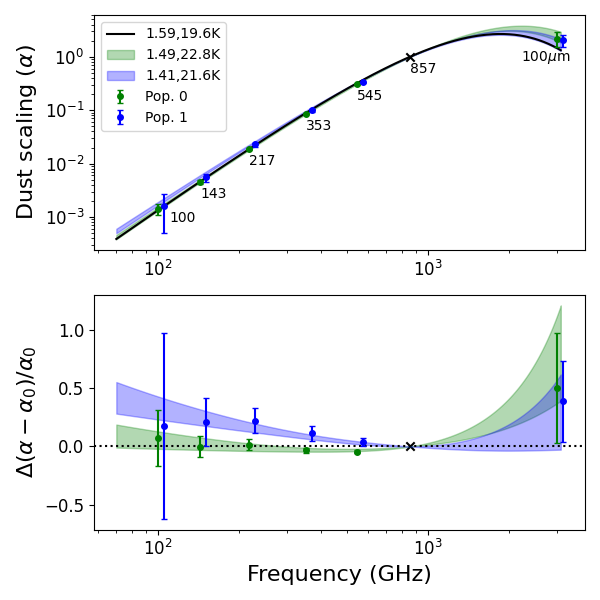}
\caption{Inferred dust properties from the two dust populations as shown in Figure \ref{fig:planck_double_mask}. The mask chosen covers the cleanest 40\% sky fraction, constructed from 857$\times$353\,GHz, applied to all frequencies \Planck HFI and \IRAS 100\,$\mu\mathrm{m}$, and only high confidence pixels were included in the analysis with $\geq 3\sigma$ confidence in assignment of dust populations. On the left, we have the best-fit dust parameters $(\beta_d, T_d)$. Plotted in black is \Planck's 2015 thermal dust model from the Commander pipeline: \footnote{\lstinline{COM_CompMap_dust-commander_0256_R2.00.fits}} distribution of $(\beta_d, T_d)$ values masked to the same region. The two populations here have overlapping contours only at the $\sim 2.5\sigma$ with the alternate population (in blue) at fixed $\beta_d$, preferring a colder $T_d$, or at fixed $T_d$, preferring a shallower $\beta_d$. This indicates that there are physical differences between the two populations. On the right, the SED model is compared to the measured flux ratios with their full bandpass corrections. The alternate population, the blue points, are shifted in frequency purely for visual separation of the two populations. The SED model accurately captures the measurements at all frequencies.} 
\label{fig:dust_inference}
\end{figure*}

Having identified the two dust populations, we can further infer the properties of each. 
We construct a mask to optimally separate the two populations. 
We use a 857$\times$353\,GHz $\sigma$-map, as shown in Figure \ref{fig:planck_double_mask}, and include only high-confidence pixels with $\geq 3\sigma$ confidence in the assignment of the dust population. 
We apply this mask to all frequencies, \Planck HFI and \IRAS 100\,$\mu$m. 
Then, we subtract a previously fitted monopole correction term and calculate the flux ratio $\alpha$ relative to our reference frequency for each frequency and population. 

The $\alpha$'s are shown on the right side of Figure \ref{fig:dust_inference}, where the uncertainties signify a $1\sigma$ variation in the population. These populations are inherently non-Gaussian. To analyze them, we perform a Gaussian kernel density estimate (KDE), denoted by $\alpha_{\textrm{kde}}$, using Scott's rule for bandwidth estimation \citep{scotts_rule, scipy}. 

The flux ratio can be related to a standard dust SED by the following definition:  
\begin{equation}
  \alpha(\nu, \beta_d, T_d) = \frac{I(\nu)}{I(\nu_0)} = \frac{b_c(\nu)}{b_c(\nu_0)}\left( \frac{\nu}{\nu_0} \right)^{\beta_d} \frac{B(\nu, T_d)}{B(\nu_0, T_d)},
  \label{eq:dust_alpha}
\end{equation}
where $\beta_d$ is the spectral index of dust, $B(\nu, T_d)$ is the Planck blackbody emission at temperature $T_d$, $\nu_0$ is the reference frequency 857\,GHz, and $b_c(\nu)$ is a bandpass correction for map $\nu$ as defined by the reciprocal of Equation 35 in \citep{planck2013_spectralresponse}. The full spectral bandpass for \Planck \footnote{\lstinline{HFI_RIMO_R3.00.fits}} and \IRAS \footnote{\lstinline{Table II. C.5 of IRAS Explanatory Supplement}} are both used for the color correction. Fitting flux ratios instead of fluxes directly allows us to avoid fitting for the dust optical depth, as it is not a fundamental property of dust physics. 

We construct a likelihood function based on the Gaussian KDE of $\alpha$,
\begin{equation} 
\log \mathcal{L}(\beta_d, T_d) = \sum_\nu^{ \text{ HFI, \IRAS 100}} \log \alpha_{\mathrm{kde}}(\alpha(\nu, \beta_d, T_d)).
\label{eq:likelihood_alpha}
\end{equation}
To find the best-fit dust parameters, we maximize the likelihood with respect to $\beta_d, T_d$. We choose uniform flat priors for $\beta_d, T_d$. The fits are performed with 
\begin{lstinline}e eemcee\end{lstinline}\citep{emcee} and are shown on the left of Figure \ref{fig:dust_inference}. We see distinctive dust properties, with the majority of the dust being a ``hot'' dust (plotted in green), with interspersed ``colder'' dust (plotted in blue).

The marginal parameters show substantial overlap between the two dust populations because of the high degeneracy between $\beta_d-T_d$. 
Intuitively, this degeneracy naturally occurs because both $\beta_d$ and $T_d$ can influence the peak of the blackbody function. An underestimation of $\beta_d$ leads to a more emissive Rayleigh-Jeans portion of the curve and pulls the peak towards lower frequencies. To compensate, a hotter dust temperature pushes the peak back towards higher frequencies \citep{Chen_2016}. 
This anti-correlation has been observed in other analyses \citep{planck_XIV_2014}. \cite{Shetty_2009} has attributed this anti-correlation to measurement uncertainty. While a hierarchical Bayesian model has been demonstrated to mitigate this effect \citep{Kelly_2012}, this modeling extension was not pursued in this analysis. 

The 2D likelihood contours show separation of the two distributions at the mutual $\sim$$2.5\sigma$ level. These values of $\beta_d, T_d$ accurately reproduce the observed flux ratios at all frequencies, as evident on the right side of Figure \ref{fig:dust_inference}. This implies that we have distinct spatially separated dust populations, with each population being effectively characterized by a single-component SED. 
It is important to remember that the uncertainties in these parameter estimations represent the variability of the populations rather than statistical uncertainty on the populations themselves.

\section{Conclusion}
\label{sec:conclusion}

We have developed a powerful technique for spatially separating dust populations based on Gaussian mixture models using HMC methods. We constructed a GMM likelihood function with no inherent spatial information. The analysis results provide, for each pixel $i$, a probability $\pi_{i,k}$ that it belongs to population $k$. 
We have shown it is highly accurate at recovering input parameters, even in the limit where the dust populations are heavily mixed.

We then applied this method to the \Planck HFI intensity maps to identify pixels that statistically deviate from a main population. 
When we show the arrangement of pixels, coherent structures emerge, showing distinctive regions where an alternate population is favored. 

These large-scale SED variations are consistent with those found in \cite{Delouis_2021}, despite employing significantly different analysis methodology. 
Furthermore, this analysis notably diminishes the role of noise in the identifications of SED variations, leading to the identification of distinctive and coherent structures that may belong to different populations. 

We tested this conclusion to many different analysis choices: (1) in Section \ref{sec:lr40_analysis}, we used entirely independent frequency pairs, i.e., 545$\times$217\,GHz and 857$\times$353\,GHz, and found consistent results, (2) in Section \ref{sec:spider_analysis}, we changed the masked area and showed that multiple dust populations are detectable even in cleaner sky regions with more diffuse dust, (3) in Section \ref{sec:dust_novar}, we explored the sensitivity of the likelihood model ($V_k = 0$) and confirmed that this effect is still observed when the likelihood is modified. Then, we probed if these effects can be sourced from solar dipole residuals. (4) in Section \ref{sec:filter}, we filtered the maps remove residual large-scale artifacts and confirmed that this effect is not attributable to such artifacts, and (5) in Section \ref{sec:bracket}, we added back a dipole and showed the identified dust populations remain stable even with residual artifacts twice the anticipated size.

Then, in Section \ref{sec:evidence}, we calculate the Bayesian evidence to see which of these analysis choices are favored. The most favored model for the majority of frequency pairs is a two-population dust model with inherent variability of each population $V_k > 0$. As the signal to noise on dust drops, so does the Bayesian evidence for this effect. Indeed, for the 545$\times$217 pair, we compute the highest Bayesian evidence for a three-dust population with no variability ($V_k = 0$). 

We also provide a promising path for improving this analysis. All datasets could be combined and simultaneously fit with a multi-variate likelihood. We expect such an extension to be computationally feasible since the dominant contributor to variable count does not scale with the number of input maps. Additionally, in this work, we excluded spatial information in the likelihood in due caution to be unbiased in dust population recovery. Combining statistical power of neighboring pixels can provide additional statistical weight that is otherwise unused in this analysis, potentially revealing further insights about the variability of dust. 

Our findings suggest that a spatially varying two-population dust model better aligns with the observed data, over a one-population model with SED properties that must vary widely to capture the variable properties across the sky.
Our analysis distinguishes between them solely based on SED scalings and cannot provide insights into their underlying compositions. The populations could signify two distinct dust components with different chemical compositions, or may originate from a one-component model experiencing vastly different optical environments, resulting in different dust temperatures. 
Identifying the source of these SED variations, particularly if stemming from two dust components, would offer deeper insights into the nature of Galactic dust. 

Nevertheless, with a two-population dust model, CMB component separation becomes a substantially more challenging task. 
Failure to correctly account for the spectral variations can lead to residual contamination.
To effectively distinguish these dust populations, we need high-fidelity, high-frequency measurements.
Indeed, this analysis would not have been possible without using intensity measurements from \Planck 545 and 857\,GHz. 

A similar understanding of dust polarization remains an open question. Polarized maps do not exist at comparable high frequencies.
Further insights into the nature of polarized dust emission necessitate higher sensitivity data, particularly in the sub-mm and far-infrared ranges. 
These measurements are accessible only through next-generation sub-orbital and orbital experiments. 

In conclusion, this analysis is another step in understanding the complexity of Galactic foreground emission. The improvement of statistical tools, in conjunction with data from \Planck's highest frequency channels, has enabled us to differentiate between multiple un-polarized dust populations. Increased instrumental sensitivity of next-generation CMB experiments will lead to data that is more discerning. Simple foreground models may no longer suffice. A continued refinement of foreground models is essential to enhance our ability to extract ever-tighter cosmological constraints from the CMB.

\begin{acknowledgements}

This research is supported by the National Aeronautics and Space Administration grant  NNX17AC55G. 

We want to express our thanks to Elle Shaw. Her careful review and helpful editorial suggestions significantly improved our work. Many of the results in this paper have been derived using the HEALPix package \citep{HealPix}. Computations were performed on the Niagara supercomputer at the SciNet HPC Consortium. SciNet is funded by Innovation, Science and Economic Development Canada; the Digital Research Alliance of Canada; the Ontario Research Fund: Research Excellence; and the University of Toronto \citep{Niagara}.

\end{acknowledgements}

\FloatBarrier

\appendix
\section{Derivation of the Mahalanobis regression distance and angle}

\begin{figure}[h]

\centering
\begin{tikzpicture}[scale = 0.75]
\draw[->, thick](0,0) to (0, 5) node[left]{$y$}; 
\node(point) at (4,0.5){};
\draw(point)[color = blue!10, fill = blue!10]circle[x radius = 1em, y radius = 4em, rotate = 0];
\draw(point)[color = blue!30, fill = blue!30]circle[x radius = 0.5em, y radius = 2em, rotate = 0];
\draw(point) [color = black, fill = black]circle(.4ex);
\draw (point) node[below]{${\bf Z_i}$};


\draw(point)[color=black] to (3.1, 2.3); 
\draw(3.2, 2.1)[black] to (3.0,2.0)[black] to (2.9, 2.2); 
\draw(3.4, 1.5) node[below, black]{$\Delta_i$};

\draw(3.7, 2.6)[dashed, color = black] to (5.3, 2.6); 
\draw(point)[color = blue, thick] to (3.7,2.6); 
\draw(3.57, 2.5) node[below, blue]{${\bf u}$};  

\draw[blue](3.95, 2.65) node[thick, below]{$\phi$}; 
\draw[thick, blue, <->](4.25, 2.6) arc(0:-65:0.75);

\draw(4.7, 2.55) node[thick, above]{$\theta$}; 
\draw[black, <->](5.0, 2.6) arc(0:26.5:1.3); 

\draw[<->, thick, red](-1.5,  0)to (0, 0.75)to (5.5, 3.5);  
\draw[red](0,0) -- (0, 0.75);
\draw[red](0, 0.375) node[left]{$b$}; 
\draw[red](0.7, 0.92) node[]{$\theta$};

\draw[dashed, black](0, 0.75) to (2, 0.75); 
\draw[->, black](0, 0.75) to (-0.25, 1.25) node[left]{${\bf v}$};
\draw[black, thin](-0.1, 0.95) to (0.1, 1.05) to(0.2, 0.85); 

\draw[->, thick](0,0) to (5,0) node[below]{$x$};
\end{tikzpicture}

 \caption{Diagram fitting a point ${\bf Z_i}$ with covariance ${\bf \Sigma_i}$ to a model line $(\theta, b)$. A mirrored geometry of Figure \ref{fig:orth_reg} showing equivalent expressions. }
 
 \label{fig:orth_reg2}
\end{figure}
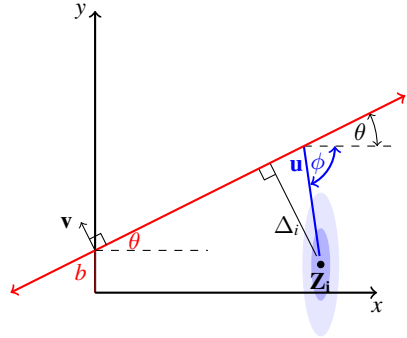

We want to derive the expression that minimizes the Mahalanobis distance between a data point with $mu_0 = (x_0, y_0)^\mathrm{T}$ and covariance $\Sigma$ and a line parametrized by $X = (x,y)^{\mathrm{T}} = (x, mx + b)^{\mathrm{T}}$. The Mahalanobis distance is defined, 
\begin{equation} 
d^2 = (X - \mu_0)^\mathrm{T} \Sigma^{-1} (X - \mu_0).
\end{equation}
We assume no off-diagonal terms are in the covariance matrix as frequency maps are independent.  Then this expression is simply, 
\begin{equation} 
d^2 = \frac{(x - x_0)^2}{\sigma_x^2} + \frac{(y - y_0)^2}{\sigma_y^2}.
\end{equation}
We want to find a point $(\hat{x}, \hat{y})$ along the line $y = mx + b$ such that the Mahalanobis distance is minimized. Taking the derivative of the distance squared and setting to zero, 
\begin{align*} 
\left. \frac{\partial d^2}{\partial x} \right|_{x = \hat{x}} &= \frac{2(\hat{x} - x_0)}{\sigma_x^2} + \frac{2m(\hat{y}- y_0)}{\sigma_y^2} = 0 \\
(\hat{x} - x_0) \sigma_y^2 &= -m \sigma_x^2 (\hat{y} - y_0) \\ 
\hat{x} \sigma_y^2 + m \sigma_x^2 \hat{y} &= x_0 \sigma_y^2 + m y_0 \sigma_x^2 \\ 
\hat{x} (\sigma_y^2 + m^2 \sigma_x^2) + mb\sigma_x^2 &= \\ 
\hat{x} &= \frac{x_0 \sigma_y^2 + m(y_0  - b)\sigma_x^2  }{\sigma_y^2 + m^2 \sigma_x^2}.
\end{align*}
Then, using the relationship that $y =mx + b$, 
\begin{align*} 
\hat{y} &= \frac{mx_0 \sigma_y^2 + m^2(y_0  - b)\sigma_x^2  + b(\sigma_y^2 + m^2 \sigma_x^2)}{\sigma_y^2 + m^2 \sigma_x^2}  \\ 
&= \frac{(mx_0 + b)\sigma_y^2 + m^2y_0 \sigma_x^2  }{\sigma_y^2 + m^2 \sigma_x^2}. \\ 
\end{align*}
Then, we can find the angle at which minimizes the Mahalanobis distance, 
\begin{align*} 
\phi &= \arctan \left( \frac{y_0 - \hat{y} }{x_0 - \hat{x}} \right) \\
&= \arctan \left(\frac{y_0 \sigma_y^2 + y_0 m^2 \sigma_x^2 - (m x_0 + b) \sigma_y^2 - m^2 y_0 \sigma_x^2}{x_0 \sigma_y^2 + x_0 m^2 \sigma_x^2 - x_0 \sigma_y^2 - m (y_0 - b) \sigma_x^2} \right) \\ 
&= \arctan \left( \frac{y_0 \sigma_y^2 - (mx_0 + b) \sigma_y^2}{x_0 m^2 \sigma_x^2 - m(y_0 - b) \sigma_x^2}\right) \\ 
&= \arctan \left( -\frac{1}{m} \frac{\sigma_y^2}{\sigma_x^2}  \frac{y_0 -  (mx_0 + b)}{ -x_0 m + (y_0 - b)}\right) \\ 
&= \arctan \left( - \frac{1}{m} \frac{\sigma_y^2}{\sigma_x^2} \right).
\end{align*}
The corresponding distance from point $(x_0, y_0)$ to the line is then found by trigonometry,
\begin{equation}
d = \frac{\Delta}{\cos(|\phi| + \theta - \pi)}.
\end{equation}
It is important to remember that the geometry for computing the distance $d$, as shown in fig \ref{fig:orth_reg} and \ref{fig:orth_reg2}, already takes into account the sign of the angle. Therefore, it would be inappropriate to use \lstinline{arctan2} in the calculation of $\phi$. 

\label{sec:appendix_derivation}
\FloatBarrier
\bibliography{references}{}
\bibliographystyle{aasjournal}

\end{document}